\documentclass[journal]{IEEEtran}

\begin{document}
\title{Fuzzy Discrete Event Systems under Fuzzy Observability and a Test-Algorithm}

\author{Daowen~Qiu and Fuchun~Liu% <-this % stops a space
\thanks{This work was supported in part by the National
Natural Science Foundation under Grant 90303024 and Grant
60573006, and the Research Foundation for the Doctorial Program of
Higher School of Ministry of Education  under Grant 20050558015,
of China.}% <-this % stops a space
\thanks{D.W. Qiu is with Department of
Computer Science, Zhongshan University, Guangzhou, 510275, China
(e-mail: issqdw@mail.sysu.edu.cn).}% <-this % stops a space
\thanks{F.C. Liu is with  1) Department
of Computer Science, Zhongshan University, Guangzhou, 510275,
China; 2) Faculty of Computer, Guangdong University of Technology,
Guangzhou, 510090,
China  (e-mail: liufch@gdut.edu.cn).}% <-this % stops a space
\thanks{}}


\markboth{IEEE TRANSACTIONS ON FUZZY SYSTEMS,~Vol.~, No.~, ~} {QIU
AND LIU: \MakeLowercase{\textit{}}: Fuzzy Discrete Event Systems
under Fuzzy Observability and a Test-Algorithm}

\maketitle

\begin{abstract}
%\boldmath
In order to more effectively cope with the real-world problems of
vagueness, impreciseness, and subjectivity, fuzzy discrete event
systems (FDESs) were proposed recently. Notably, FDESs have been
applied to biomedical control for HIV/AIDS treatment planning and
sensory information processing for robotic control. Qiu, Cao and
Ying independently developed supervisory control theory of FDESs.
We note that the controllability of events in Qiu's work is fuzzy
but the observability of events is crisp, and, the observability
of events in Cao and Ying's work is also crisp although the
controllability is not completely crisp since the controllable
events can be disabled with any degrees. Motivated by the
necessity to consider the situation that the events may be
observed or controlled with some membership degrees, in this
paper, we establish the supervisory control theory of FDESs with
partial observations, in which both the observability and
controllability of events are fuzzy instead. We formalize the
notions of fuzzy controllability condition and fuzzy observability
condition. And Controllability and Observability Theorem of FDESs
is set up in a more generic framework. In particular, we present a
detailed computing flow to verify whether the controllability and
observability conditions hold. Thus, this result can decide the
existence of supervisors. Also, we use this computing method to
check the existence of supervisors in the Controllability and
Observability Theorem of classical discrete event systems (DESs),
which is a new method and different from classical case. A number
of examples are elaborated on to illustrate the presented results.
\end{abstract}


\section{Introduction}
\IEEEPARstart{D}ISCRETE event systems (DESs) are dynamical systems
whose evolution in time is governed by the abrupt occurrence of
physical events at possibly irregular time intervals. Event though
DESs are quite different from traditional continuous variable
dynamical systems, they clearly involve objectives of control and
optimization. A fundamental issue of supervisory control for DESs
is how to design a controller (or supervisor), whose task is to
enable and disable the controllable events such that the resulting
closed-loop system obeys some prespecified operating rules [1]. Up
to now, the supervisory control theory of DESs has been
significantly applied to many technological and engineering
systems such as automated manufacturing systems, interaction
telecommunication networks and protocol verification in
communication networks [2-9].

In most of engineering applications, the states of a DES are
crisp. However, this is not the case in many other applications in
complex systems such as biomedical systems and economic systems,
in which vagueness, impreciseness, and subjectivity are typical
features. For example, it is vague when a man's condition of the
body is said to be ``good". Moreover, it is imprecise to say at
what point exactly a man has changed from state ``good" to state
``poor". It is well known that the fuzzy set theory first proposed
by Zadeh [10] is a good tool to cope with those problems. Indeed,
up to now, fuzzy control systems have been well developed by many
authors, and  we may refer to [11] (and these references therein)
regarding a survey on model-based fuzzy control systems. Notably,
 Lin and Ying [12, 13] recently initiated significantly the study of {\it
fuzzy discrete event systems} (FDESs) by combining fuzzy set
theory [14] with classical DESs. Notably, FDESs have been applied
to biomedical control for HIV/AIDS treatment planning [15, 16] and
decision making [17]. More recently, R. Huq {\it et al} [18, 19]
have proposed an intelligent sensory information processing
technique using FDESs for robotic control in the field of mobile
robot navigation.

Just as Lin and Ying [13] pointed out, a comprehensive theory of
FDESs still needs to be set up, including many important concepts,
methods and theorems, such as controllability, observability, and
optimal control. These issues have been partially investigated in
[20-23]. It is worthy to mention that Qiu [20], Cao and Ying [21]
independently developed the supervisory control theory of FDESs.
The similarity between the two theories is that the fuzzy systems
considered in both [20] and [21] are modeled by max-min automata
instead of max-product automata adopted in [13], and the
controllability theorem was established in their respective
frameworks. However, there are great differences between them. For
the purpose of control, the set of events in [21] is partitioned
into two disjoint subsets of controllable and uncontrollable
events, as usually done in classical DESs, but the controllability
of events is not completely crisp since the controllable events
can be disabled by supervisors with any degrees. In contrast with
[21], the controllable set and uncontrollable set of events in
[20] are two {\it fuzzy subsets} of the set of events. That is,
each event not only belongs to the uncontrollable set but also
belongs to the controllable set; only its degree of belonging to
those sets may be different. In particular, Qiu [20] presented an
algorithm to check the existence of fuzzy supervisors for FDESs.
As a continuation of the supervisory control under full
observations [20, 21], this paper is to deal with the supervisory
control of FDESs with fuzzy observations (generalizing partial
observations).

We notice that the observability in Qiu's work [20] and Cao and
Ying's work [21-23] is {\it crisp}, that is, each fuzzy event is
either completely observable or completely unobservable, although
the controllability is fuzzy in [20] and not completely crisp in
[21-23] where the controllable events can be disabled with any
degrees. However, in real-life situation, each event generally has
a certain degree to be observable and unobservable, and, also, has
a certain degree to be controllable and uncontrollable. In fact,
this idea of fuzziness of observability and controllability was
originally proposed by Lin and Ying [13], and Qiu [20], and then
it has been subsequently applied to robot sensory information
processing by Huq {\it et al} [18, 19]. For example, in the cure
process for a patient having cancer via either operation or drug
therapy [24], some treatments (events) can be clearly seen by
supervisors (viewed as a group of physicians), while some
therapies (such as some operations) may not completely be observed
by supervisors. For another example, in order to provide
state-based decision making for a physical agent in mobile robot
control, Huq {\it et al} [18, 19] introduced the concept of
state-based observability to interpret the degree of reliability
of the sensory information used in constructing fuzzy event
matrices.

Motivated by the necessity to consider the situation that the
events may be observed or controlled with some membership degrees,
in this paper, we establish the supervisory control theory of
FDESs with partial observations, in which both the observability
and the controllability of events are {\it fuzzy} instead. We
formalize the notions of fuzzy controllability condition and fuzzy
observability condition. A Controllability and Observability
Theorem of FDESs is set up in a more generic framework. In
particular, we present a computing flow to verify whether the
controllability and observability conditions hold, which can
decide the existence of supervisors. Also, we apply this computing
method to testing the existence of supervisors in the
Controllability and Observability Theorem of classical DESs [1],
which is a different method from classical case [1].

The remainder of the paper is organized as follows. In the
interest of readability, in Section II, we recall related notation
and notions in supervisory control theory of FDESs. In Section
III, we establish a Controllability and Observability Theorem of
FDESs. Section IV deals with the realization of supervisors in the
theorem; we present a computing flow for testing the existence of
supervisors. Also, we elaborate on a number of related examples to
illustrate the presented results.


% An example of a floating table. Note that, for IEEE style tables, the
% \caption command should come BEFORE the table. Table text will default to
% \footnotesize as IEEE normally uses this smaller font for tables.
% The \label must come after \caption as always.
%
%\begin{table}[!t]
%% increase table row spacing, adjust to taste
%\renewcommand{\arraystretch}{1.3}
% if using array.sty, it might be a good idea to tweak the value of
% \extrarowheight as needed to properly center the text within the cells
%\caption{An Example of a Table}
%\label{table_example}
%\centering
%\begin{tabular}{|c||c|}
%\hline
%One & Two\\
%\hline
%Three & Four\\
%\hline
%\end{tabular}
%\end{table}

% Note that IEEE does not put floats in the very first column - or typically
% anywhere on the first page for that matter. Also, in-text middle ("here")
% positioning is not used. Most IEEE journals use top floats exclusively.
% Note that, LaTeX2e, unlike IEEE journals, places footnotes above bottom
% floats. This can be corrected via the \fnbelowfloat command of the
% stfloats package.

\section{Preliminaries}
Firstly we give some notation. ${\cal P}(X)$ denotes the power set
of set $X$. A fuzzy subset of set $X$ is defined as a mapping from
$X$ to $ [0,1]$. The set of all fuzzy subsets over $X$ is denoted
as ${\cal F}(X)$. For two fuzzy subsets $\widetilde{A}$ and
$\widetilde{B}$, $\widetilde{A}\subseteq \widetilde{B}$ stands for
$\widetilde{A}(x) \leq \widetilde{B}(x)$ for any element $x$ of
domain.

A nondeterministic finite automaton [25] is a system described by
$ G=(Q,E,\delta,q_{0},Q_{m})$, where $Q$ is the finite set of
states with the initial state $q_{0}$, $E$ is the finite set of
events, $\delta: Q\times E\rightarrow {\cal P}(Q)$ is the
transition relation, and $Q_{m}\subseteq Q$ is called the set of
marked states. Each sequence over $E$ is called a {\it string}.
$E^{*}$ denotes the set of all finite strings over $E$. For $u\in
E^{*}$, $|u|$ denotes the length of $u$; if $|u|=0$, then $u$ is
an empty string, denoted by $\epsilon$. A subset of $E^{*}$ is
called a {\it language}.

In the setting of FDESs, states are fuzzy subsets of the crisp
state set $Q$, which are called {\it fuzzy states}. If the crisp
state set $Q=\{q_0, q_1,\dots, q_{n-1}\}$, then each fuzzy state
$\widetilde{q}$ can be written as a vector $[a_{0}\hskip 1mm
a_{1}\hskip 1mm\cdots\hskip 1mm a_{n-1}]$, where $a_{i}\in[0,1]$
represents the possibility of the current state being $q_{i}$.
Similarly, a {\it fuzzy event} $\widetilde{\sigma}$ is denoted by
a matrix $[a_{ij}]_{n\times n}$, in which every entry $a_{ij}$
belongs to $[0,1]$ and means the possibility of system
transforming from the current state $q_{i}$ to state $q_{j}$ when
event $\sigma$ occurs.

{\it Definition 1:} A fuzzy finite automaton is a max-min system
$$\widetilde{G}=(\widetilde{Q},\widetilde{E},\widetilde{\delta},\widetilde{q}_{0},
\widetilde{Q}_{m}),$$ where $\widetilde{Q}$ is a set of fuzzy
states; $\widetilde{E}$ consists of fuzzy events;
$\widetilde{q}_{0}$ is the initial state;
$\widetilde{Q}_m\subseteq \widetilde{Q}$ is the set of marking
states; the state transition relation $\widetilde{\delta}:
\widetilde{Q}\times \widetilde{E}\rightarrow \widetilde{Q}$ is
defined as $\widetilde{\delta}(\widetilde{q},\widetilde{\sigma})=
\widetilde{q}\odot \widetilde{\sigma}$. Note that $\odot$ is {\it
max-min operation} introduced in fuzzy set theory [26]: for matrix
$A=[a_{ij}]_{n\times m}$ and matrix $B=[b_{ij}]_{m\times k}$,
define $A\odot B=[c_{ij}]_{n\times k}$, where
$c_{ij}=\max_{l=1}^{m} \min\{a_{il}, b_{lj}\}$.

The fuzzy languages generated and marked by $\widetilde{G}$,
denoted by ${\cal L}_{\widetilde{G}}$ and ${\cal
L}_{\widetilde{G},m}$, respectively, are defined as two functions
from $\widetilde{E}^{*}$  to [0,1] as follows: for any $
\widetilde{\sigma}_{1}\cdots\widetilde{\sigma}_{k} \in
\widetilde{E}^{*}$,
\begin{equation}
{\cal L}_{\widetilde{G}}
(\widetilde{\sigma}_{1}\cdots\widetilde{\sigma}_{k})
=\max_{i=1}^{n}\widetilde{q}_{0}\odot
\widetilde{\sigma}_{1}\odot\cdots\odot\widetilde{\sigma}_{k}\odot
\overline{s}_{i}^{{\rm T}},
\end{equation}
\begin{equation}
{\cal
L}_{\widetilde{G},m}(\widetilde{\sigma}_{1}\cdots\widetilde{\sigma}_{k})
=\max_{\widetilde{q}\in \widetilde{Q}_{m}} \widetilde{q}_{0}\odot
\widetilde{\sigma}_{1}\odot\cdots\odot\widetilde{\sigma}_{k}\odot
\widetilde{q}^{{\rm T}},
\end{equation}
where ${\rm T}$ is transpose operation and
$\overline{s}_{i}=[0\cdots1\cdots0]$ where 1 is in the $i$th
place. The following property is obtained in [20]: for any $
\widetilde{s} \in \widetilde{E}^{*}$ and any $ \widetilde{\sigma}
\in \widetilde{E}$,
\begin{equation}
{\cal L}_{\widetilde{G},m}( \widetilde{s}  \widetilde{\sigma}
)\leq {\cal L}_{\widetilde{G}}( \widetilde{s}  \widetilde{\sigma}
)\leq {\cal L}_{\widetilde{G}}( \widetilde{s} ).
\end{equation}

{\it Remark 1:} The framework of this paper is based on [20-23] in
which the set of fuzzy events is a finite set. Indeed, the above
definition of fuzzy finite automaton is similar to the fuzzy
automaton defined by Steimann and Adlassning [27] for dealing with
an application of clinical monitoring. Furthermore, we would like
to consider max-min automata usually in practical applications
since the set of fuzzy states $\{\widetilde{q}_{0}\odot
\widetilde{s}:\widetilde{s}\in \widetilde{E}^{*}\}$ in any max-min
automaton is clearly finite [27]. For fuzzy automata theory and
related applications, we can refer to [28-31].

We further need some notions. A {\it sublanguage} of ${\cal
L}_{\widetilde{G}}$ is represented as $\widetilde{K}\in {\cal
F}(\widetilde{E})$ satisfying $\widetilde{K}\subseteq {\cal
L}_{\widetilde{G}}$. For $\widetilde{s}\in\widetilde{E}^{*}$,
symbol $pr(\widetilde{s})$ represents all of the prefix substrings
of $\widetilde{s}$. And for any fuzzy language ${\cal L}$ over
$\widetilde{E}$, its prefix-closure fuzzy language $pr({\cal
L}):\widetilde{E}^{*}\rightarrow [0,1]$ is defined as
\[
pr({\cal L})(\widetilde{s})=\sup_{\widetilde{s}\in
pr(\widetilde{t})}{\cal L}(\widetilde{t}),
\]
which denotes the possibility of string $\widetilde{s}$ belonging
to the prefix-closure of ${\cal L}$.

\section{Controllability and Observability Theorem}
Let
$\widetilde{G}=(\widetilde{Q},\widetilde{E},\widetilde{\delta},\widetilde{q}_{0},
\widetilde{Q}_{m})$ be a fuzzy finite automaton. As mentioned in
Section I, each fuzzy event may be observable or controllable with
a certain membership degree. Thus, the uncontrollable set
$\widetilde{\Sigma}_{uc}$ and controllable set
$\widetilde{\Sigma}_{c}$, as well as, the unobservable set
$\widetilde{\Sigma}_{uo}$ and observable set
$\widetilde{\Sigma}_{o}$, are thought of as four fuzzy subsets of
$\widetilde{E}$, which are defined formally as follows.

{\it Definition 2:} The uncontrollable set
$\widetilde{\Sigma}_{uc}\in {\cal F}(\widetilde{E})$ and
controllable set $\widetilde{\Sigma}_{c}\in {\cal
F}(\widetilde{E})$ are respectively defined as a function
$\widetilde{\Sigma}_{uc}: \widetilde{E}\rightarrow [0,1]$ and a
function $\widetilde{\Sigma}_{c}: \widetilde{E}\rightarrow [0,1]$
which satisfy: for any $\widetilde{\sigma} \in \widetilde{E}$,
\begin{equation}
\widetilde{\Sigma}_{uc}(\widetilde{\sigma})+\widetilde{\Sigma}_{c}(\widetilde{\sigma})=1.
\end{equation}
Similarly, the unobservable set $\widetilde{\Sigma}_{uo}\in {\cal
F}(\widetilde{E})$ and observable set $\widetilde{\Sigma}_{o}\in
{\cal F}(\widetilde{E})$ are respectively defined as
$\widetilde{\Sigma}_{uo}: \widetilde{E}\rightarrow [0,1]$ and
$\widetilde{\Sigma}_{o}: \widetilde{E}\rightarrow [0,1]$ which
satisfy: for any $\widetilde{\sigma} \in \widetilde{E}$,
\begin{equation}
\widetilde{\Sigma}_{uo}(\widetilde{\sigma})+\widetilde{\Sigma}_{o}(\widetilde{\sigma})=1.
\end{equation}

{\it Remark 2:} The degrees of observability and unobservability
for FDESs were originally proposed by Lin and Ying ([13], pp.
412), and the degrees of controllability and uncontrollability
were introduced by Qiu ([20], pp. 76). Intuitively,
$\widetilde{\Sigma}_{uc}(\widetilde{\sigma})$ and
$\widetilde{\Sigma}_{c}(\widetilde{\sigma})$ represent the degree
of fuzzy event $\widetilde{\sigma}$ to be uncontrollable and the
degree of $\widetilde{\sigma}$ to be controllable, respectively.
And, $\widetilde{\Sigma}_{uo}(\widetilde{\sigma})$ and
$\widetilde{\Sigma}_{o}(\widetilde{\sigma})$ represent the degree
of $\widetilde{\sigma}$ to be unobservable and the degree of
$\widetilde{\sigma}$ to be observable, respectively.

{\it Definition 3:} The projection $P:\widetilde{E}\rightarrow
\widetilde{E}$ is defined as:
\begin{equation}
P(\widetilde{\sigma})=\left\{\begin{array}{ll}
\widetilde{\sigma},& {\rm if}\hskip 2mm \widetilde{\Sigma}_{o}(\widetilde{\sigma})>0,\\
\epsilon, & {\rm otherwise}.
\end{array}\right.
\end{equation}
And it can be extended to $\widetilde{E}^{*}$ by
$P(\epsilon)=\epsilon$ and
$P(\widetilde{s}\widetilde{\sigma})=P(\widetilde{s})
P(\widetilde{\sigma})$ for $\widetilde{s}\in\widetilde{E}^{*}$ and
$\widetilde{\sigma}\in \widetilde{E}$.

{\it Remark 3:} The purpose of projection is to erase the
completely unobservable fuzzy events in the strings.

In order to emphasize the observability degree of fuzzy event
strings by means of projection $P$ associated with the fuzzy
observable subset $\widetilde{\Sigma}_{o}$, we define the {\it
factor of observable projection} $\widetilde{D}$ as a fuzzy subset
of $P(\widetilde{E})$: for any $\widetilde{\sigma}\in
P(\widetilde{E})$,
$\widetilde{D}(\widetilde{\sigma})=\widetilde{\Sigma}_{o}(\widetilde{\sigma})$,
and $$\widetilde{D}(\widetilde{\sigma}_{1}\widetilde{\sigma}_{2}
\cdots\widetilde{\sigma}_{n})=
\min\{\widetilde{D}(\widetilde{\sigma}_{i}): i=1,2,\cdots,n\},$$
where $\widetilde{\sigma}_{i}\in P(\widetilde{E})$,
$i=1,2,\cdots,n$. Especially, $\widetilde{D}(\epsilon)=0$.
Intuitively, $\widetilde{D}(P(\widetilde{s}))\cdot{\cal
L}_{\widetilde{G}}(\widetilde{s})$ represents the possibility for
the fuzzy event string $\widetilde{s}\in\widetilde{E}^{*}$ being
possible under the effect of observable projection. And,
$\widetilde{D}(P(\widetilde{s}\widetilde{\sigma}))\cdot\widetilde{\Sigma}_{uc}(\widetilde{\sigma})$
and $\widetilde{D}(P(\widetilde{s}))\cdot
pr(\widetilde{K})(\widetilde{s})$ respectively denote the degree
of $\widetilde{\sigma}\in\widetilde{E}$, as a continuation of the
string $\widetilde{s}$, being uncontrollable, and the possibility
of string $\widetilde{s}$ belonging to the prefix-closure of
sublanguage $\widetilde{K}$ under the effect of observable
projection. Furthermore, for the sake of convenience, in what
follows we use the following notation:
\begin{equation}
{\cal
L}_{\widetilde{G}}^{f}(\widetilde{s})=\left\{\begin{array}{ll}
1,& {\rm if}\hskip 2mm \widetilde{s}=\epsilon,\\
\widetilde{D}(P(\widetilde{s}))\cdot{\cal
L}_{\widetilde{G}}(\widetilde{s}), & {\rm otherwise};
\end{array}\right.
\end{equation}
\begin{equation}
pr(\widetilde{K})^{f}(\widetilde{s})=\left\{\begin{array}{ll}
1,& {\rm if}\hskip 2mm \widetilde{s}=\epsilon,\\
\widetilde{D}(P(\widetilde{s}))\cdot
pr(\widetilde{K})(\widetilde{s}), & {\rm otherwise};
\end{array}\right.
\end{equation}
\begin{equation}
\widetilde{\Sigma}_{uc}^{f}(\widetilde{\sigma})
=\widetilde{D}(P(\widetilde{s}\widetilde{\sigma}))
\cdot\widetilde{\Sigma}_{uc}(\widetilde{\sigma}),
\end{equation}
where $\widetilde{\sigma}$ is the continuation of string
$\widetilde{s}$.

{\it Definition 4:} For any FDES $\widetilde{G}$, a supervisor
under the projection $P$ is said a {\it fuzzy supervisor}, denoted
by $\widetilde{S}_{P}$, that is formally defined as a function
$$\widetilde{S}_{P}:P(\widetilde{E}^{*})\rightarrow
{\cal F}(\widetilde{E})$$ where for each
$\widetilde{s}\in\widetilde{E}^{*}$ and $\widetilde{\sigma}\in
\widetilde{E}$, \hskip 3mm
$\widetilde{S}_{P}(P(\widetilde{s}))(\widetilde{\sigma})$
represents the possibility of fuzzy event $\widetilde{\sigma}$
being enabled after the occurrence of the string
$P(\widetilde{s})$.

The supervisors $\widetilde{S}_{P}$ are usually required to
satisfy the following admissibility condition.

{\it Definition 5:} The {\it fuzzy admissibility condition} for
fuzzy supervisor $\widetilde{S}_{P}$ is characterized as follows:
for each $\widetilde{s}\in\widetilde{E}^{*}$ and each continuation
$\widetilde{\sigma}\in \widetilde{E}$, the following inequality
holds
\begin{equation}
\min\{\widetilde{\Sigma}_{uc}^{f}(\widetilde{\sigma}),\hskip
2mm{\cal
L}_{\widetilde{G}}^{f}(\widetilde{s}\widetilde{\sigma})\}\leq
\widetilde{S}_{P}(P(\widetilde{s}))(\widetilde{\sigma}).
\end{equation}

Intuitively, the fuzzy admissibility condition (10) means that,
under the effect of observable projection, the degree of any fuzzy
event $\widetilde{\sigma}$ following any fuzzy event string
$\widetilde{s}$ being possible together with $\widetilde{\sigma}$
being uncontrollable is not larger than the possibility for
$\widetilde{\sigma}$ being enabled by the fuzzy supervisor
$\widetilde{S}_{P}$ after string $P(\widetilde{s})$ occurring.

The fuzzy controlled system by means of $\widetilde{S}_{P}$,
denoted by $ \widetilde{S}_{P}/\widetilde{G} $, is an FDES, and,
the behavior of $ \widetilde{S}_{P}/\widetilde{G} $ when
$\widetilde{S}_{P}$ is controlling $\widetilde{G}$ is defined as
follows.

{\it Definition 6:} The fuzzy languages $ {\cal
L}_{\widetilde{S}_{P}/\widetilde{G}} $ and ${\cal
L}_{\widetilde{S}_{P}/\widetilde{G},m}$ generated and marked by
$\widetilde{S}_{P}/\widetilde{G}$, respectively, are defined as
follows: for any $\widetilde{s}\in\widetilde{E}^{*}$ and any
$\widetilde{\sigma}\in \widetilde{E}$,

1) $ {\cal L}_{\widetilde{S}_{P}/\widetilde{G}}(\epsilon)=1$;

2) $ {\cal
L}_{\widetilde{S}_{P}/\widetilde{G}}(\widetilde{s}\widetilde{\sigma})=
\min\{{\cal L}_{\widetilde{S}_{P}/\widetilde{G}}(\widetilde{s}),
\hskip 1mm{\cal
L}_{\widetilde{G}}^{f}(\widetilde{s}\widetilde{\sigma}),\hskip
1mm\widetilde{S}_{P}(P(\widetilde{s}))(\widetilde{\sigma})\}$;

3) $ {\cal L}_{\widetilde{S}_{P}/\widetilde{G},m} ={\cal
L}_{\widetilde{S}_{P}/\widetilde{G}} \widetilde{\cap} {\cal
L}_{\widetilde{G},m}$,\\
 where symbol $\widetilde{\cap}$ denotes Zadeh fuzzy AND operator, i.e.,
$(\widetilde{A}\widetilde{\cap}
\widetilde{B})(x)=\min\{\widetilde{A}(x),\widetilde{B}(x)\}$.

Definition 6 indicates that the degree of
$\widetilde{s}\widetilde{\sigma}$ being physically possible in the
controlled system $\widetilde{S}_{P}/\widetilde{G}$ is the
smallest one among the degree of $\widetilde{s}$ being possible in
$\widetilde{S}_{P}/\widetilde{G}$, the degree of
$\widetilde{s}\widetilde{\sigma}$ being possible in
$\widetilde{G}$ under the effect of observable projection, and the
possibility of $\widetilde{\sigma}$ being enabled by the
supervisor after the occurrence of $P(\widetilde{s})$. It is clear
that Definition 6 generalizes the corresponding concepts from full
observations ([20], pp. 6) to partial observations.

In supervisory control of DESs, nonblockingness is usually
required, and it means that the controlled system does not produce
deadlocks [1, 20].

{\it Definition 7:} A fuzzy supervisor $\widetilde{S}_{P}$ of
$\widetilde{G}$ is said to be {\it nonblocking}, if for any
$\widetilde{s}\in\widetilde{E}^{*}$, the following equation holds:
\begin{equation}
{\cal
L}_{\widetilde{S}_{P}/\widetilde{G}}(\widetilde{s})=\left\{\begin{array}{ll}
1,& {\rm if}\hskip 2mm \widetilde{s}=\epsilon,\\
\widetilde{D}(P(\widetilde{s}))\cdot pr({\cal
L}_{\widetilde{S}_{P}/\widetilde{G},m})(\widetilde{s}), & {\rm
otherwise}.
\end{array}\right.
\end{equation}

Intuitively, if $\widetilde{S}_{P}$ is nonblocking, then for any
string $\widetilde{s}$, the possibility that $\widetilde{s}$ is
one of the behaviors of the supervised fuzzy system
$\widetilde{S}_{P}/\widetilde{G}$ equals the degree of
$\widetilde{s}$ belonging to the prefix-closure of the fuzzy
language marked by the supervised fuzzy system
$\widetilde{S}_{P}/\widetilde{G}$ under the effect of observable
projection.

{\it Definition 8:} A fuzzy sublanguage $\widetilde{K}$ is said to
be {\it ${\cal L}_{\widetilde{G},m}$-closed}, if for any
$\widetilde{s}\in\widetilde{E}^{*}$,
\begin{equation}
\widetilde{K}(\widetilde{s})=\left\{\begin{array}{ll}
1,& {\rm if}\hskip 2mm \widetilde{s}=\epsilon,\\
\min\{pr(\widetilde{K})^{f}(\widetilde{s}), \hskip 1mm{\cal
L}_{\widetilde{G},m}(\widetilde{s})\}, & {\rm otherwise}.
\end{array}\right.
\end{equation}

Obviously, if all fuzzy events can be observed fully [20], that is
to say, $\widetilde{\Sigma}_{o}(\widetilde{\sigma})=1$ for any
fuzzy event $\widetilde{\sigma}$, then Eq. (12) reduces to
$K=pr(K)$ introduced in [1, 4, 7, 20], where all events are
supposed to be observable.

{\it Definition 9:} Let $\widetilde{K}\subseteq {\cal
L}_{\widetilde{G}}$. If for any
$\widetilde{s}\in\widetilde{E}^{*}$ and its continuation
$\widetilde{\sigma}\in\widetilde{E}$, the following inequality
holds:
\begin{eqnarray}
\min\{pr(K)^{f}(\widetilde{s}),\hskip 1mm
\widetilde{\Sigma}_{uc}^{f}(\widetilde{\sigma}),\hskip 1mm {\cal
L}_{\widetilde{G}}^{f}(\widetilde{s}\widetilde{\sigma})\}\leq
pr(\widetilde{K})^{f}(\widetilde{s}\widetilde{\sigma}),
\end{eqnarray}
then we call $\widetilde{K}$ satisfying {\it fuzzy controllability
condition with respect to $\widetilde{G}$, $P$ and
$\widetilde{\Sigma}_{uc}$}.

Intuitively, (13) means that under the effect of observable
projection, the degree to which any fuzzy event string
$\widetilde{s}$ belongs to the prefix-closure of $\widetilde{K}$
and fuzzy event $\widetilde{\sigma}$ following string
$\widetilde{s}$ is physically possible together with
$\widetilde{\sigma}$ being uncontrollable, is not larger than the
possibility of string $\widetilde{s}\widetilde{\sigma}$ belonging
to the prefix-closure of $\widetilde{K}$.

{\it Remark 4:} Definition 9 generalizes the corresponding
concepts concerning controllability in [1, 20]. If all fuzzy
events can be observed fully, then Ineq. (13) reduces to the fuzzy
controllability condition introduced in [20]. If we further assume
that the events and states are crisp, then it reduces to the
controllability condition introduced in [1].

To illustrate the application of fuzzy controllability condition,
we provide an example.

{\it Example 1.} Consider a fuzzy automaton
$\widetilde{G}=(\widetilde{Q}_{1},\widetilde{E},\widetilde{\delta},
\widetilde{q}_{0})$, where
$\widetilde{E}=\{\widetilde{a},\widetilde{b}, \widetilde{c}\}$,
$\widetilde{q}_{0}$=$[0.8, 0]$, and
\[
\widetilde{a}=\left[
\begin{array}{cc}
0.8 &\ 0.2\\
0&\ 0.2
\end{array}
\right], \widetilde{b}=\left[
\begin{array}{cc}
0.2 &\ 0.8\\
0&\ 0.2
\end{array}
\right], \widetilde{c}=\left[
\begin{array}{cc}
0.2&\ 0\\
0.8&\ 0.2
\end{array}
\right].
\]
Let $pr(\widetilde{K})$ be generated by a fuzzy automaton
$\widetilde{H}=(\widetilde{Q}_{2},\widetilde{E},\widetilde{\delta},\widetilde{p}_{0})$,
where $\widetilde{p}_{0}$=$[0.5,0]$,
$\widetilde{E}=\{\widetilde{a},\widetilde{b}, \widetilde{c}\}$,
and $\widetilde{a},\widetilde{b}$ are the same as those in
$\widetilde{G}$ , but $\widetilde{c}$ is changed as follows:
\[
\widetilde{c}=\left[
\begin{array}{cc}
0.1&\ 0\\
0.4&\ 0.1
\end{array}
\right].
\]
Suppose that $\widetilde{\Sigma}_{uc}$ and
$\widetilde{\Sigma}_{o}$ are defined as:
$\widetilde{\Sigma}_{uc}(\widetilde{a})=0.3$,\hskip 2mm
$\widetilde{\Sigma}_{uc}(\widetilde{b})=0.5$, and
$\widetilde{\Sigma}_{uc}(\widetilde{c})= 0.8$;\hskip 2mm
$\widetilde{\Sigma}_{o}(\widetilde{a})=\widetilde{\Sigma}_{o}(\widetilde{b})=0.7$,
 and $\widetilde{\Sigma}_{o}(\widetilde{c})= 0.5$.

In the following, we show that $\widetilde{K}$ is not fuzzy
controllable. Take $\widetilde{s}=\widetilde{b}$ and $
\widetilde{\sigma}=\widetilde{c}$. Then
\begin{eqnarray*}
&&\min\{pr(K)^{f}(\widetilde{s}),\hskip 2mm
\widetilde{\Sigma}_{uc}^{f}(\widetilde{\sigma}),\hskip 2mm {\cal
L}_{\widetilde{G}}^{f}(\widetilde{s}\widetilde{\sigma})\}\\
&=&\min\left\{0.7\times0.5,\hskip 1mm 0.5\times0.8, \hskip
2mm0.5\times0.8\right\}=0.35.
\end{eqnarray*}
However,
$$pr(\widetilde{K})^{f}(\widetilde{s}\widetilde{\sigma})
=0.5\times0.4=0.2.$$ Therefore, the fuzzy controllability
condition does not hold.

If $\widetilde{\Sigma}_{uc}$ is changed into
$\widetilde{\Sigma}_{uc}(\widetilde{\sigma})\leq 0.05$ for any
$\widetilde{\sigma}\in \widetilde{E}$, then we can check that the
fuzzy controllability condition holds.

Before setting up the Controllability and Observability Theorem of
FDESs, we need a characterization of the observability of fuzzy
sublanguage.

 {\it Definition 10:} Let $\widetilde{K}\subseteq {\cal
L}_{\widetilde{G}}$. If for any
$\widetilde{s}\in\widetilde{E}^{*}$
 and $\widetilde{\sigma}\in\widetilde{E}$, the following inequality
holds:
\begin{eqnarray}
\min\{pr(K)^{f}(\widetilde{s}),
pr(\widetilde{K})^{f}(\widetilde{t}\widetilde{\sigma}), {\cal
L}_{\widetilde{G}}^{f}(\widetilde{s}\widetilde{\sigma})\}\leq
pr(\widetilde{K})^{f}(\widetilde{s}\widetilde{\sigma})
\end{eqnarray}
for any $\widetilde{t}\in \widetilde{\Sigma}^{*}$, where
$P(\widetilde{s})=P(\widetilde{t})$, then $\widetilde{K}$ is said
satisfying {\it fuzzy observability condition with respect to
$\widetilde{G}$ and $P$}.

Intuitively, (14) means that if there is another string
$\widetilde{t}$ possessing the same projection as $\widetilde{s}$,
then under the effect of observable projection, the degree to
which string $\widetilde{s}$ belongs to the prefix-closure of
$\widetilde{K}$ and fuzzy event $\widetilde{\sigma}$ following
$\widetilde{s}$ is physically possible together with
$\widetilde{t}\widetilde{\sigma}$ belonging to the prefix-closure
of $\widetilde{K}$, is not larger than the possibility of string
$\widetilde{s}\widetilde{\sigma}$ belonging to the prefix-closure
of $\widetilde{K}$.

{\it Example 2.}  Consider a fuzzy automaton
$\widetilde{G}=(\widetilde{Q}_{1},\widetilde{E},
\widetilde{\delta},\widetilde{q}_{0})$, where
$\widetilde{E}=\{\widetilde{a},\widetilde{b}, \widetilde{c},
\widetilde{d}\}$, $\widetilde{q}_{0}$=$[0.8, 0]$, and
\[
\widetilde{a}=\left[
\begin{array}{cc}
0.8 &\ 0.2\\
0&\ 0.2
\end{array}
\right], \hskip 3mm \widetilde{b}=\left[
\begin{array}{cc}
0.2 &\ 0.8\\
0&\ 0.2
\end{array}
\right], \]\[ \widetilde{c}=\left[
\begin{array}{cc}
0.5 &\ 0\\
0.4&\ 0.5
\end{array}
\right], \hskip 3mm \widetilde{d}=\left[
\begin{array}{cc}
0.2&\ 0\\
0.8&\ 0.2
\end{array}
\right].
\]
Let $pr(\widetilde{K})$ be generated by
$\widetilde{H}=(\widetilde{Q}_{2},\widetilde{E},\widetilde{\delta},\widetilde{p}_{0})$,
where $\widetilde{p}_{0}$=$[0.5,0]$,
$\widetilde{E}=\{\widetilde{a},\widetilde{b},
\widetilde{c},\widetilde{d}\}$, and $\widetilde{a},\widetilde{b}$
are the same as those in $\widetilde{G}$ , but $\widetilde{c}$ and
$\widetilde{d}$ are changed as follows:
\[
\widetilde{c}=\left[
\begin{array}{cc}
0.3 &\ 0\\
0.4&\ 0.3
\end{array}
\right], \hskip 3mm \widetilde{d}=\left[
\begin{array}{cc}
0.2&\ 0\\
0.4&\ 0.2
\end{array}
\right].
\]
Suppose that $\widetilde{\Sigma}_{o}$ is defined as:
$\widetilde{\Sigma}_{o}(\widetilde{a})=0.5$,\hskip 2mm
$\widetilde{\Sigma}_{o}(\widetilde{b})=0.7$, \hskip 2mm
$\widetilde{\Sigma}_{o}(\widetilde{c})= 0.4$, and
$\widetilde{\Sigma}_{o}(\widetilde{d})=0$.

If we take $\widetilde{s}=\widetilde{b}\widetilde{d}$, $
\widetilde{\sigma}=\widetilde{c}$ and
$\widetilde{t}=\widetilde{b}$, then
$P(\widetilde{s})=P(\widetilde{t})$, and
\begin{eqnarray*}
&&\min\left\{pr(K)^{f}(\widetilde{s}),\hskip 2mm
pr(\widetilde{K})^{f}(\widetilde{t}\widetilde{\sigma}),\hskip 2mm
{\cal
L}_{\widetilde{G}}^{f}(\widetilde{s}\widetilde{\sigma})\right\}\\
&=&\min\left\{0.7\times0.4,\hskip 2mm 0.4\times0.4, \hskip
2mm0.4\times0.5\right\}=0.16.
\end{eqnarray*}
However,
$$pr(\widetilde{K})^{f}(\widetilde{s}\widetilde{\sigma})
=0.4\times0.3=0.12.$$ Therefore, the fuzzy observability condition
does not hold.

On the basis of the preliminaries, we are ready to present the
main theorem of the paper.

{\it Theorem 1:} ({\it Controllability and Observability Theorem
of FDESs}).  Let $\widetilde{G}=(\widetilde{Q},\widetilde{E},
\widetilde{\delta},\widetilde{q}_{0},\widetilde{Q}_{m})$ be a
fuzzy automaton with a projection $P$. Suppose that fuzzy language
$\widetilde{K}\subseteq {\cal L}_{\widetilde{G},m}$ satisfies
$\widetilde{K}(\epsilon)=1$ and $pr(\widetilde{K})\subseteq {\cal
L}_{\widetilde{G},m}$. Then there exists a nonblocking fuzzy
supervisor $\widetilde{S}_{P}:P(\widetilde{E}^{*})\rightarrow
{\cal F}(\widetilde{E})$, such that $\widetilde{S}_{P}$ satisfies
the fuzzy admissibility condition, and
\[
{\cal L}_{\widetilde{S}_{P}/\widetilde{G}}(\widetilde{s})=
pr(\widetilde{K})^{f}(\widetilde{s}) \hskip 7mm {\rm and} \hskip
7mm {\cal L}_{\widetilde{S}_{P}/\widetilde{G},m}(\widetilde{s})=
\widetilde{K}(\widetilde{s})
\]
for any $\widetilde{s}\in \widetilde{E}^{*}$, if, and only if the following conditions hold:\\
1. $\widetilde{K}$ satisfies fuzzy controllability condition
w.r.t. $\widetilde{G}$, $P$ and
$\widetilde{\Sigma}_{uc}$. \\
2. $\widetilde{K}$ satisfies fuzzy observability condition w.r.t. $\widetilde{G}$ and $P$.\\
3. $\widetilde{K}$ is ${\cal L}_{\widetilde{G},m}$-closed.

\begin{proof} See Appendix.\end{proof}

\section{Realization of Supervisors in Controllability
and Observability Theorem of FDESs}
In this section, we present a
detailed computing method to verify the controllability and
observability conditions. Thus, this method can decide the
existence of supervisors in Controllability and Observability
Theorem of FDESs. As applications, two examples are elaborated to
illustrate that this computing method is suitable to check the
existence of supervisors not only for FDESs but also for classical
DESs.

\subsection{Method of Checking the Existence of Supervisors for FDESs}
Clearly, the existence of supervisor is associated with both fuzzy
controllability condition and fuzzy observability condition.
Therefore, testing the two conditions described by Ineqs. (13, 14)
is of great importance. In classical DESs, for a given automaton
$G$ and a language $K$, the controllability condition is checked
by comparing the active event set of each state of $H\times G$
with the active event set of each state of $G$, where automaton
$H$ generates $pr(K)$. And the observability condition is checked
by building an observer of an automaton with unobservable events
at each site [1].

For FDESs, a computing method of checking the fuzzy
controllability condition was given by Qiu [20]. Based on the main
idea of the finiteness of fuzzy states in FDESs modeled by max-min
automata, we present a detailed approach for testing the fuzzy
observability condition by means of {\it computing trees}. Let
$\widetilde{G}=(\widetilde{Q},\widetilde{E},\widetilde{\delta},
\widetilde{q}_{0},\widetilde{Q}_{m})$ be a fuzzy automaton with
partial observations and
$\widetilde{E}=\{\widetilde{a}_{1},\widetilde{a}_{2},\cdots,\widetilde{a}_{n}\}$.
Assume that the prefix-closure of fuzzy language
$\widetilde{K}\subseteq {\cal L}_{\widetilde{G},m}$ is generated
by a fuzzy automaton
$\widetilde{H}=(\widetilde{Q}_{1},\widetilde{E},\widetilde{\delta},\widetilde{p}_{0})$.
We describe the computing process via three steps as follows.

The first step gives a computing tree for deriving the set of all
fuzzy states reachable from the initial state $\widetilde{q}_{0}$,
and the sets of strings respectively corresponding to each
accessible fuzzy state are also obtained. The basic idea is based
on the following two points:
\begin{itemize}
\item $\widetilde{p}_{0}\odot \widetilde{s}=\widetilde{p}_{0}\odot
\widetilde{s}\odot (\widetilde{t})^{k}$ for any $k\geq 0$ if
$\widetilde{p}_{0}\odot \widetilde{s}=\widetilde{p}_{0}\odot
\widetilde{s}\odot \widetilde{t}$ for $ \widetilde{t}\in
\widetilde{E}^{*}$, where $(\widetilde{t})^{k}$ denotes the
$\odot$ product of $k$'s $\widetilde{t}$. \item The set of fuzzy
states $\{\widetilde{p}_{0}\odot \widetilde{s}: s\in
\widetilde{E}^{*}\}$ is always finite since $\widetilde{E}$ is
finite [20]. \end{itemize}

Without loss of generality, we present the computing tree for
$\widetilde{E}=\{\widetilde{a}_{1},\widetilde{a}_{2}\}$ of two
fuzzy events via Fig. 1, and the case of more than two fuzzy
events is analogous.

{\it Step 1:} For a fuzzy automaton
$\widetilde{H}=(\widetilde{Q}_{1},\widetilde{E},\widetilde{\delta},\widetilde{p}_{0})$,
we search for all possible fuzzy states $\widetilde{r}_{i}$
reachable from $\widetilde{p}_{0}$ in $\widetilde{H}$,
$i=1,2,\ldots,m_{1}$; also, we can obtain the sets
$C(\widetilde{r}_{i})$ of all fuzzy event strings whose inputs
lead $\widetilde{p}_{0}$ to $\widetilde{r}_{i}$,
$i=1,2,\ldots,m_{1}$. This process can be realized by the finite
computing tree that is visualized by Fig. 1 as follows.

In the computing tree, the initial fuzzy state $\widetilde{p}_{0}$
is its root; each vertex, say $\widetilde{p}_{0}\odot
\widetilde{s}$, may produce $n$'s sons, i.e.,
$\widetilde{p}_{0}\odot \widetilde{s}\odot \widetilde{a}_{i}$,
$i=1,2,\ldots,n$. However, if $\widetilde{p}_{0}\odot
\widetilde{s}\odot \widetilde{a}_{i}$ equals some its father, then
$\widetilde{p}_{0}\odot \widetilde{s}\odot \widetilde{a}_{i}$ is a
leaf, that is marked by a underline. The computing ends with a
leaf at the end of each branch.

\setlength{\unitlength}{0.05in}

\begin{picture}(60,47)

\put(25,40){\makebox(10,5)[c]{{\tiny Begin}}}
\put(30,41){\vector(0,-1){2}} \put(25,34){\makebox(10,5)[c]{{\tiny
$\widetilde{p}_{0}$}}} \put(30,34){\line(0,-1){2}}
\put(25,32){\line(1,0){10}}
\put(25,32){\vector(0,-1){5}}\put(35,32){\vector(0,-1){5}}

\put(20,22){\makebox(10,5)[c]{{\tiny
$\widetilde{p}_{0}\odot\widetilde{a}_1$}}}\put(30,22){\makebox(10,5)[c]{{\tiny
$\widetilde{p}_{0}\odot\widetilde{a}_2$}}}
\put(25,22){\vector(0,-1){8}}\put(8,19){\line(1,0){17}}
\put(8,19){\vector(0,-1){5}}
\put(35,22){\vector(0,-1){8}}\put(35,19){\line(1,0){17}}
\put(52,19){\vector(0,-1){5}} \put(2,9){\makebox(10,5)[c]{{\tiny
$\widetilde{p}_{0}\odot\widetilde{a}_1\odot\widetilde{a}_1$}}}
\put(20,9){\makebox(10,5)[c]{{\tiny
$\widetilde{p}_{0}\odot\widetilde{a}_1\odot\widetilde{a}_2$}}}
\put(34,9){\makebox(10,5)[c]{{\tiny
$\widetilde{p}_{0}\odot\widetilde{a}_2\odot\widetilde{a}_1$}}}
\put(50,9){\makebox(10,5)[c]{{\tiny
$\widetilde{p}_{0}\odot\widetilde{a}_2\odot\widetilde{a}_2$}}}

\put(4,6){\makebox(10,5)[c]{{\vdots}}}
\put(20,6){\makebox(10,5)[c]{{\vdots}}}
\put(33,6){\makebox(10,5)[c]{{\vdots}}}
\put(47,6){\makebox(10,5)[c]{{\vdots}}}

\put(20,27){\makebox(5,5)[c]{{\tiny
$\widetilde{a}_1$}}}\put(35,22){\makebox(5,15)[c]{{\tiny
$\widetilde{a}_2$}}} \put(3,9){\makebox(5,15)[c]{{\tiny
$\widetilde{a}_1$}}}\put(20,9){\makebox(5,15)[c]{{\tiny
$\widetilde{a}_2$}}} \put(35,9){\makebox(5,15)[c]{{\tiny
$\widetilde{a}_1$}}}\put(52,9){\makebox(5,15)[c]{{\tiny
$\widetilde{a}_2$}}}

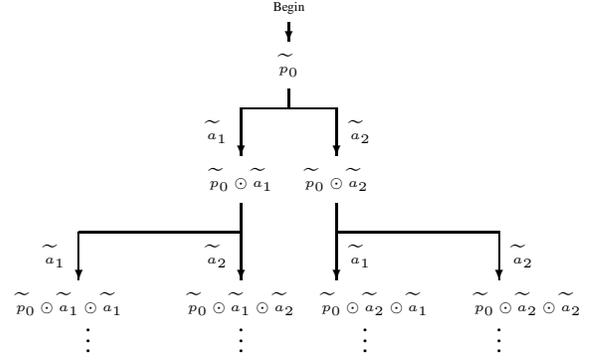
\put(23,0){\makebox(20,5)[c]{{\footnotesize Fig. 1. Computing tree
of all states reachable from $\widetilde{p}_{0}$. }}}

\end{picture}

For two fuzzy automata $\widetilde{G}$ and $\widetilde{H}$, our
purpose is to search for the all different fuzzy state pairs
reachable from the initial fuzzy state pair
$(\widetilde{q}_{0},\hskip 2mm\widetilde{p}_{0})$. The method is
similar to Step 1, which is also carried out by a computing tree.
In this computing tree, the root is labelled with pair
$(\widetilde{q}_{0},\hskip 1mm\widetilde{p}_{0})$, and each
vertex, say $(\widetilde{q}_{0}\odot \widetilde{s},\hskip
2mm\widetilde{p}_{0}\odot \widetilde{s})$ for $\widetilde{s}\in
\widetilde{E}^{*}$, may produce $n$'s sons, i.e.,
$(\widetilde{q}_{0}\odot \widetilde{s}\odot
\widetilde{a}_{i},\hskip 2mm\widetilde{p}_{0}\odot
\widetilde{s}\odot \widetilde{a}_{i})$, $i=1,2,\ldots,n$. But if a
pair $(\widetilde{q}_{0}\odot \widetilde{s}\odot
\widetilde{a}_{i},\hskip 2mm\widetilde{p}_{0}\odot
\widetilde{s}\odot \widetilde{a}_{i})$ is the same as one of its
fathers, then this pair will be treated as a leaf, that is marked
with a underline. Such a computing tree is depicted by Fig. 2.
Since the set of all fuzzy state pairs is finite, the computing
tree ends with a leaf at the end of each branch.

{\it Step 2:} For fuzzy automata $\widetilde{G}$ and
$\widetilde{H}$, we search for all possible pairs of fuzzy states
$(\widetilde{q}_{i},\widetilde{p}_{i})$, $i=1,2,\ldots,m_{2}$,
reachable from $(\widetilde{q}_{0},\widetilde{p}_{0})$ by a finite
computing tree (Fig. 2), and, in the same time, we can decide the
sets $C(\widetilde{q}_{i},\widetilde{p}_{i})$ of all fuzzy event
strings each of which makes
$(\widetilde{q}_{0},\widetilde{p}_{0})$ become
$(\widetilde{q}_{i},\widetilde{p}_{i})$, $i=1,2,\ldots,m_{2}$.

\setlength{\unitlength}{0.05in}

\begin{picture}(60,58)

\put(28,50){\makebox(10,5)[c]{{\tiny Begin}}}
 \put(33,51){\vector(0,-1){3}}

 \put(28,44){\makebox(10,5)[c]{{\tiny
($\widetilde{q}_{0}$,\hskip 2mm $\widetilde{p}_{0}$)}}}

\put(33,44){\line(0,-1){3}} \put(23,41){\line(1,0){20}}
\put(23,41){\vector(0,-1){5}}\put(43,41){\vector(0,-1){5}}
\put(13,31){\makebox(10,5)[c]{{\tiny ($\widetilde{q}_{0}\odot
\widetilde{a}_1$,\hskip 2mm
$\widetilde{p}_{0}\odot\widetilde{a}_1$)}
}}\put(43,31){\makebox(10,5)[c]{{\tiny
($\widetilde{q}_{0}\odot\widetilde{a}_2$,\hskip 2mm
   $\widetilde{p}_{0}\odot\widetilde{a}_2$)}}}
\put(23,31){\vector(0,-1){16}} \put(6,28){\line(1,0){17}}
\put(6,28){\vector(0,-1){5}} \put(43,31){\vector(0,-1){8}}
\put(43,28){\line(1,0){15}} \put(58,28){\vector(0,-1){13}}

 \put(3,18){\makebox(10,5)[c]{{\tiny
($\widetilde{q}_{0}\odot\widetilde{a}_1^{2}$,\hskip 2mm
$\widetilde{p}_{0}\odot\widetilde{a}_1^{2}$)}
}}\put(17,10){\makebox(10,5)[c]{{\tiny
($\widetilde{q}_{0}\odot\widetilde{a}_1\odot\widetilde{a}_2$,\hskip
2mm
$\widetilde{p}_{0}\odot\widetilde{a}_1\odot\widetilde{a}_2$)}}}
\put(38,18){\makebox(10,5)[c]{{\tiny
($\widetilde{q}_{0}\odot\widetilde{a}_2\odot\widetilde{a}_1$,\hskip
2mm $\widetilde{p}_{0}\odot\widetilde{a}_2\odot\widetilde{a}_1$)}
}}\put(51,10){\makebox(10,5)[c]{{\tiny
($\widetilde{q}_{0}\odot\widetilde{a}_2^{2}$,\hskip 2mm
$\widetilde{p}_{0}\odot\widetilde{a}_2^{2}$)}}}
\put(18,36){\makebox(5,5)[c]{{\tiny
$\widetilde{a}_1$}}}\put(38,36){\makebox(5,5)[c]{{\tiny
$\widetilde{a}_2$}}} \put(1,23){\makebox(5,5)[c]{{\tiny
$\widetilde{a}_1$}}}\put(23,23){\makebox(5,5)[c]{{\tiny
$\widetilde{a}_2$}}} \put(38,23){\makebox(5,5)[c]{{\tiny
$\widetilde{a}_1$}}}\put(52,23){\makebox(5,5)[c]{{\tiny
$\widetilde{a}_2$}}}

\put(1,14){\makebox(10,5)[c]{{\vdots}}}
\put(16,7){\makebox(10,5)[c]{{\vdots}}}
\put(38,14){\makebox(10,5)[c]{{\vdots}}}
 \put(53,7){\makebox(10,5)[c]{{\vdots}}}

\put(23,1){\makebox(20,5)[c]{{\footnotesize Fig. 2. Computing tree
of all state pairs reachable from
$(\widetilde{q}_{0},\widetilde{p}_{0})$.} }}
\end{picture}

We now present Step 3, and, following that, we will give a
proposition to further show the feasibility of this step.

{\it Step 3:} Set
$P(\widetilde{q}_{i},\widetilde{p}_{i})=\{\widetilde{s^{'}}|
P(\widetilde{s^{'}})=P(\widetilde{s}),\widetilde{s}\in
C(\widetilde{q}_{i},\widetilde{p}_{i})\}$, $i=1,2,\ldots,m_{2}$,
and further set
\begin{equation}
R_{j}(\widetilde{q}_{i},\widetilde{p}_{i})=\{\widetilde{t}|
P(\widetilde{q}_{i},\widetilde{p}_{i})\cap
C(\widetilde{r}_{j})\not=\emptyset, \widetilde{p}_{0}\odot
\widetilde{t}=\widetilde{r}_{j}\},
\end{equation}
for $i=1,2,\ldots,m_{2}$, and $j=1,2,\ldots,m_{1}$. If
$R_{j}(\widetilde{q}_{i},\widetilde{p}_{i})\not=\emptyset$, we
arbitrarily choose a string, say $\widetilde{t}_{ij}\in
R_{j}(\widetilde{q}_{i},\widetilde{p}_{i})$ (usually, we try to
choose a shorter string, and this will decrease our computing
complexity in what follows). Given any $i\in\{1,2,\ldots,m_{2}\}$,
by $FR(\widetilde{q}_{i},\widetilde{p}_{i})$ we mean the set of
all strings $\widetilde{t}_{ij}$ we have chosen, say
$$FR(\widetilde{q}_{i},\widetilde{p}_{i})=
\{\widetilde{t}_{i1},\widetilde{t}_{i2},\ldots,\widetilde{t}_{ik_{i}}\}.$$
If Ineq. (14) holds for each $\widetilde{s}_{i}\in
C(\widetilde{q}_{i},\widetilde{p}_{i})$ and each
$\widetilde{t}_{ij}\in FR(\widetilde{q}_{i},\widetilde{p}_{i})$
where $i\in\{1,2,\ldots,m_{2}\}$ and $j\in\{1,2,\ldots,k_{i}\}$,
then the fuzzy observability condition (14) holds; otherwise it
does not hold. This is further verified by the following
Proposition 2.

{\it Proposition 2:}  Let
$\widetilde{G}=(\widetilde{Q},\widetilde{E},\widetilde{\delta},
\widetilde{q}_{0},\widetilde{Q}_{m})$ and
$\widetilde{H}=(\widetilde{Q}_{1},\widetilde{E},\widetilde{\delta},\widetilde{p}_{0})$
be two fuzzy automata. Suppose that fuzzy sublanguage
$\widetilde{K}$ satisfies $pr(\widetilde{K})={\cal
L}_{\widetilde{H}}\subseteq {\cal L}_{\widetilde{G},m}$. If for
any $i=1,2,\ldots,m_{2}$, there exist $\widetilde{s}_{i}\in
C(\widetilde{q}_{i},\widetilde{p}_{i})$ such that for any
$\widetilde{r}\in FR(\widetilde{q}_{i},\widetilde{p}_{i})$ and any
$\widetilde{\sigma}\in\widetilde{E}$, Ineq. (14) holds, then the
fuzzy observability condition described by Ineq. (14) holds.

\begin{proof} For any $\widetilde{t}\in\widetilde{E}^{*}$, without
loss of generality, suppose that $\widetilde{t}\in
C(\widetilde{q}_{i_{0}},\widetilde{p}_{i_{0}})$ for some $i_{0}\in
\{1,2,\ldots,m_{2}\}$, since $\widetilde{E}^{*}=
\bigcup_{i=1}^{m_{2}}C(\widetilde{q}_{i},\widetilde{p}_{i})$. For
any $\widetilde{t^{'}}\in \widetilde{E}^{*}$ satisfying
$P(\widetilde{t^{'}})=P(\widetilde{t})$, then
$\widetilde{t^{'}}\in
P(\widetilde{q}_{i_{0}},\widetilde{p}_{i_{0}})$, and we can
further assume $\widetilde{t^{'}}\in C(\widetilde{r}_{j_{0}})$ for
some $j_{0}\in \{1,2,\ldots,m_{1}\}$, due to
$\widetilde{E}^{*}=\bigcup_{j=1}^{m_{1}}C(\widetilde{r}_{j})$.
Therefore, there is $\widetilde{t}_{i_{0}j_{0}}\in
FR(\widetilde{q}_{i_{0}},\widetilde{p}_{i_{0}})$. Now we have the
following relations:
\begin{equation}
pr(\widetilde{K})(\widetilde{t})=[\widetilde{p}_{0}\odot
\widetilde{t}]=[\widetilde{p}_{0}\odot \widetilde{s}_{i_{0}}],
\end{equation}
\begin{equation}
pr(\widetilde{K})(\widetilde{t^{'}}\widetilde{\sigma})=[\widetilde{p}_{0}\odot
\widetilde{t^{'}}\odot\widetilde{\sigma}]=[\widetilde{p}_{0}\odot
\widetilde{t}_{i_{0}j_{0}}\odot\widetilde{\sigma}],
\end{equation}
\begin{equation}
pr(\widetilde{K})(\widetilde{t}\widetilde{\sigma})=[\widetilde{p}_{0}\odot
\widetilde{t}\odot\widetilde{\sigma}]=[\widetilde{p}_{0}\odot
\widetilde{s}_{i_{0}}\odot\widetilde{\sigma}],
\end{equation}
\begin{equation}
{\cal L}_{\widetilde{G}}
(\widetilde{t}\widetilde{\sigma})=[\widetilde{q}_{0}\odot
\widetilde{t}\odot\widetilde{\sigma}]=[\widetilde{q}_{0}\odot
\widetilde{s}_{i_{0}}\odot\widetilde{\sigma}].
\end{equation}
By means of the existing condition in this proposition, we know
that
\begin{eqnarray}
\begin{array}{lll}
&&\min\{pr(\widetilde{K})^{f}(\widetilde{s}_{i_{0}}),
pr(\widetilde{K})^{f}(\widetilde{t}_{i_{0}j_{0}}\widetilde{\sigma}),{\cal
L}_{\widetilde{G}}^{f}(\widetilde{s}_{i_{0}}\widetilde{\sigma})\}\\
&\leq&
pr(\widetilde{K})^{f}(\widetilde{s}_{i_{0}}\widetilde{\sigma}).
\end{array}
\end{eqnarray}
In terms of Eqs. (16-19) and Ineq. (20) we therefore obtain
\begin{eqnarray}
\begin{array}{lll}
&& \min\{pr(\widetilde{K})^{f}(\widetilde{t}),
pr(\widetilde{K})^{f}(\widetilde{t^{'}}\widetilde{\sigma}),{\cal
L}_{\widetilde{G}}^{f}(\widetilde{t}\widetilde{\sigma})\}\\
&\leq& pr(\widetilde{K})^{f}(\widetilde{t}\widetilde{\sigma}),
\end{array}
\end{eqnarray}
and this completes the proof of proposition.\end{proof}

Based on the above Proposition 2, we can check the fuzzy
observability condition described by Ineq. (14) by the above
computing flow (Steps 1--3). Furthermore, the fuzzy
controllability condition described Ineq. (13) also can be clearly
tested by similar computing flow with slight changes
($pr(\widetilde{K})(\widetilde{s}\widetilde{\sigma})$ is replaced
by $\widetilde{\Sigma}_{uc}(\widetilde{\sigma})$), besides using
the approach proposed by Qiu [20].

{\it Remark 6.} To conclude this section, we roughly analyze the
complexity of the above computing flow. Suppose that the number of
all different fuzzy states $\{\widetilde{p}_{0}\odot
\widetilde{s}: \widetilde{s}\in \widetilde{E}^{*}\}$ is $m_{1}$,
and the number of all different fuzzy state pairs reachable from
the initial fuzzy state pair
$(\widetilde{q}_{0},\widetilde{p}_{0})$, namely,
$\{(\widetilde{q}_{0}\odot \widetilde{s},\widetilde{p}_{0}\odot
\widetilde{s}): \widetilde{s}\in \widetilde{E}^{*}\}$, is $m_{2}$.
Then, in {\it Step 1}, by means of Figure 1 the number of
computing steps is $O(m_{1})$ and, in {\it Step 2}, in terms of
Figure 2 the number of computing steps is $O(m_{2})$. As to {\it
Step 3}, we can see that the computing complexity is
$O(m_{1}m_{2}|\widetilde{E}|)$, where $|\widetilde{E}|$ is the
cardinal number of alphabet $\widetilde{E}$. Thus, if we avoid the
cost regarding the operation $\odot$, the computing steps of the
above flow is $O(m_{1}m_{2}|\widetilde{E}|)$.

\subsection{Applications to Supervisory Control of Classical DESs and FDESs}
In this subsection, we present two examples to illustrate the
applications of the supervisory control theory for FDESs presented
above. Example 3 will indicate that the computing approach given
in Section IV-A can be applied to check the existence of
supervisors for classical DESs. Example 4 arising from a medical
treatment will describe a detailed computing processing for FDESs,
which may be viewed as an applicable background of supervisory
control of FDESs under partial observations.

We first recall some notions of classical DESs. Let $G$ be a
classical DES. Suppose that $\Sigma_{c}$ and $\Sigma_{o}$ are
designated as controllable and observable event sets,
respectively. $P$ is the corresponding projection. A language $K$
is said to be observable with respect to $G$ and $P$, if for all
$s$, $t\in pr(K)$ and all $\sigma\in \Sigma_{c}$, if $P(s)=P(t)$,
then
\begin{equation}
s\sigma\in {\cal L}_{G},\hskip 3mm t\sigma\in pr(K) \hskip 4mm
\Rightarrow \hskip 4mm s\sigma\in pr(K).\end{equation}

In classical DESs [1], for a given automaton $G$ and a language
$K$, the controllability condition is checked by comparing the
active event set of each state of $H\times G$ with the active
event set of each state of $G$, where automaton $H$ generates
$pr(K)$. And the observability condition is checked by building an
observer of an automaton with unobservable events at each site
[1].

{\it Example 3:} Consider the example presented in Section 3.7 of
[1] (Example 3.18, page 196) to illustrate the method of testing
the observability condition (22). $G$ and $H$ are two automata of
classical DESs with crisp state set $E=\{u, b\}$ shown in Fig. 3.
Language $K$ satisfies $pr(K)={\cal L}_{H}$. Assume that
$\Sigma_{o}=\{b\}$ and $\Sigma_{c}=\{u, b\}$. In order to test $K$
being unobservable, an observer automaton $H_{obs}$ is constructed
in [1]. In fact, the observability condition (22) cannot be
satisfied when $s=\epsilon$, $t=u$ and $\sigma=b$.

\setlength{\unitlength}{0.075cm}
\begin{picture}(150,45)

\put(10,28){\circle{5}\makebox(-10,0){$0$}}
\put(30,36){\circle{5}\makebox(-10,0){$1$}}
\put(50,28){\circle{5}\makebox(-10,0){$2$}}
\put(30,20){\circle{5}\makebox(-10,0){$3$}}

\put(2.5,28){\vector(1,0){5}}\put(12.5,29){\vector(2,1){15}}
\put(12.5,27){\vector(2,-1){15}} \put(32.5,36){\vector(2,-1){15}}
 \put(32.5,20){\vector(2,1){15}}

\put(19,35){\makebox(0,0)[c]{$u$}}
\put(19,21){\makebox(0,0)[c]{$b$}}
\put(40,35){\makebox(0,0)[c]{$b$}}
\put(40,21){\makebox(0,0)[c]{$u$}}

\put(20,11){\makebox(20,1)[c]{{\footnotesize (1) Automaton $G$}}}

\put(70,26){\circle{5}\makebox(-10,0){$0$}}
\put(87,34){\circle{5}\makebox(-10,0){$1$}}
\put(105,26){\circle{5}\makebox(-10,0){$2$}}

\put(62.5,26){\vector(1,0){5}}\put(72.5,27){\vector(2,1){12}}
\put(90,34){\vector(2,-1){13}}

\put(79,33){\makebox(0,0)[c]{$u$}}
\put(100,33){\makebox(0,0)[c]{$b$}}

\put(78,11){\makebox(20,1)[c]{{\footnotesize (2) Automaton $H$}}}
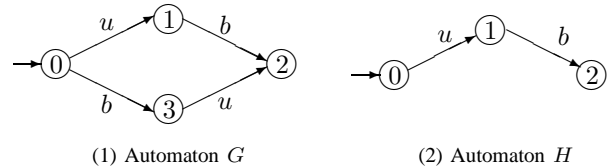
\put(50,3){\makebox(20,1)[c]{{\footnotesize  Fig. 3.  Automata $G$
and $H$ of classical DESs in Example 3.}}}

\end{picture}

In the following, we verify the above conclusion by means of the
computing method we presented in Section IV-A.

Firstly, classical DES $G$ can be viewed as a fuzzy automaton
$\widetilde{G}=(\widetilde{Q}_{1},\widetilde{E},
\widetilde{\delta},\widetilde{q}_{0})$, where the fuzzy states are
$$\widetilde{q}_{0}=[1,0,0,0],\hskip
4mm\widetilde{q}_{1}=[0,1,0,0],$$$$
\widetilde{q}_{2}=[0,0,1,0],\hskip
4mm\widetilde{q}_{3}=[0,0,0,1],$$ and the fuzzy events are
\[
\widetilde{u}=\left[
\begin{array}{cccc}
0 &\ 1 &\ 0 &\ 0\\
0 &\ 0 &\ 0 &\ 0\\
0 &\ 0 &\ 0 &\ 1\\
0 &\ 0 &\ 0 &\ 0\\
\end{array}
\right], \hskip 4mm \widetilde{b}=\left[
\begin{array}{cccc}
0 &\ 0 &\ 1 &\ 0\\
0 &\ 0 &\ 0 &\ 1\\
0 &\ 0 &\ 0 &\ 0\\
0 &\ 0 &\ 0 &\ 0\\
\end{array}
\right].
\]\\
Similarly, the automaton $H$ can be viewed as a fuzzy automaton
$\widetilde{H}=(\widetilde{Q}_{2},\widetilde{E},
\widetilde{\delta},\widetilde{p}_{0})$, where the fuzzy states are
$$\widetilde{p}_{0}=[1,0,0,0],\hskip 4mm\widetilde{p}_{1}=[0,1,0,0],\hskip 4mm
\widetilde{p}_{2}=[0,0,1,0],$$ and the fuzzy events are
\[
\widetilde{u}=\left[
\begin{array}{cccc}
0 &\ 1 &\ 0 &\ 0\\
0 &\ 0 &\ 0 &\ 0\\
0 &\ 0 &\ 0 &\ 0\\
0 &\ 0 &\ 0 &\ 0\\
\end{array}
\right], \hskip 4mm \widetilde{b}=\left[
\begin{array}{cccc}
0 &\ 0 &\ 0 &\ 0\\
0 &\ 0 &\ 0 &\ 1\\
0 &\ 0 &\ 0 &\ 0\\
0 &\ 0 &\ 0 &\ 0\\
\end{array}
\right].
\]

The fuzzy subsets $\widetilde{\Sigma}_{o}$ and
$\widetilde{\Sigma}_{c}$ are determined by $\Sigma_{o}=\{b\}$ and
$\Sigma_{c}=\{u, b\}$, which are listed as follows:
$$\widetilde{\Sigma}_{o}(\widetilde{u})=0, \hskip 4mm\widetilde{\Sigma}_{o}(\widetilde{b})=1;
\hskip
4mm\widetilde{\Sigma}_{c}(\widetilde{u})=\widetilde{\Sigma}_{c}(\widetilde{b})=1.$$

By constructing the computing trees of $\widetilde{H}$,
$\widetilde{G}$ and $\widetilde{H}$, we know that there are three
fuzzy states $\widetilde{p}_{0}$, $\widetilde{p}_{1}$,
$\widetilde{p}_{2}$ reachable from $\widetilde{p}_{0}$, and three
fuzzy states pairs $(\widetilde{q}_{0},\widetilde{p}_{0})$,
$(\widetilde{q}_{1},\widetilde{p}_{1})$,
$(\widetilde{q}_{2},\widetilde{p}_{2})$ reachable from
$(\widetilde{q}_{0},\widetilde{p}_{0})$, and the corresponding
fuzzy event strings are $\epsilon$, $\widetilde{u}$, and
$\widetilde{u}\widetilde{b}$. Therefore, we should necessarily
check the fuzzy observability condition in term of whether the all
elements in the rightmost column of the following Table I are ``T"
(True) when $\widetilde{s}=\epsilon$,
$\widetilde{s}=\widetilde{u}$, and
$\widetilde{s}=\widetilde{u}\widetilde{b}$, where
\begin{itemize}
\item $x_{1}=[\widetilde{p}_0\odot \widetilde{s}]$,
$x_{2}=[\widetilde{p}_0\odot
\widetilde{t}\odot\widetilde{\sigma}]$,
$x_{3}=[\widetilde{q}_0\odot
\widetilde{s}\odot\widetilde{\sigma}]$,\item
$y=[\widetilde{p}_0\odot
\widetilde{s}\odot\widetilde{\sigma}]$,\item
$V=\min\{pr(K)^{f}(\widetilde{s}),
pr(\widetilde{K})^{f}(\widetilde{t}\widetilde{\sigma}), {\cal
L}_{\widetilde{G}}^{f}(\widetilde{s}\widetilde{\sigma})\}$, \item
$W=pr(K)^{f}(\widetilde{s}\widetilde{\sigma})$.
\end{itemize}

\begin{table}
\caption{Testing the fuzzy observability condition in Example 3}
\begin{center}
\begin {tabular}{|c|c|c|c|c|c|c|c|c|c|}
\hline
$\widetilde{s}$&$\widetilde{t}$&$\widetilde{\sigma}$&$x_{1}$
&$x_{2}$& $x_{3}$& $y$& $V$&$ W $&
$V\leq W $\\

\hline
&$\epsilon$&$\widetilde{u}$&1&1&1&1&0&0& T \\
\cline{3-10}
$\epsilon$&&$\widetilde{b}$&1&0&1&0&0&0& T \\
\cline{2-10}
&$\widetilde{u}$&$\widetilde{u}$&1&0&1&1&0&0& T \\
\cline{3-10}
&&$\widetilde{b}$&1&1&1&0&1&0& F \\
\hline
&$\epsilon$&$\widetilde{u}$&1&1&0&0&0&0& T  \\
\cline{3-10}
$\widetilde{u}$&&$\widetilde{b}$&1&0&1&1&0&1& T \\
\cline{2-10}
&$\widetilde{u}$&$\widetilde{u}$&1&0&0&0&0&0& T  \\
\cline{3-10}
&&$\widetilde{b}$&1&1&1&1&0&1& T \\
\hline
&$\widetilde{b}$&$\widetilde{u}$&1&0&0&0&0&0& T  \\
\cline{3-10}
&&$\widetilde{b}$&1&0&0&0&0&0& T \\
\cline{2-10}
$\widetilde{u}\widetilde{b}$&$\widetilde{b}\widetilde{u}$&$\widetilde{u}$&1&0&0&0&0&0& T  \\
\cline{3-10}
&&$\widetilde{b}$&1&0&0&0&0&0& T \\
\cline{2-10}
&$\widetilde{u}\widetilde{b}$&$\widetilde{u}$&1&0&0&0&0&0& T  \\
\cline{3-10}
&&$\widetilde{b}$&1&0&0&0&0&0& T \\
\hline
\end{tabular}
\end{center}
\end{table}

From Table I we see that the fuzzy observability condition does
not hold since an ``F" (False) has been found out in the rightmost
column when $\widetilde{s}=\epsilon$,
$\widetilde{t}=\widetilde{u}$ and
$\widetilde{\sigma}=\widetilde{b}$.

Example 3 indicates that our method also can be applied to testing
the existence of supervisors for classical DESs [1]. Next we apply
our results to an applicable example arising from a medical
treatment problem.

{\it Example 4:} Suppose that there is a patient sickening for a
new disease. For simplicity, it is assumed that the doctors
consider roughly the patient's condition to be two states, say
``poor" and ``good". For the new disease, the doctors have no
complete knowledge about it, but they believe by their experience
that these drugs such as theophylline, Erythromycin Ethylsuccinate
and dopamine may be useful to the disease.

As mentioned in Introduction, considering the features of
vagueness, patient's condition can simultaneously belong to
``poor" and ``good" with respective memberships; also, an event
occurring (i.e., treatment) may lead a state to multistates with
respective degrees. Therefore, the patient's conditions and their
changes after the treatments can be modeled by an FDES
$\widetilde{G}=(\widetilde{Q}_{1},\widetilde{E},
\widetilde{\delta},\widetilde{q}_{0},\widetilde{Q}_{m})$, in which
each fuzzy state, denoted as a two-dimensional vector
$\widetilde{q}=[a_{1}, a_{2}]$, is represented as the possibility
distribution of the patient's condition over the two crisp states
``poor" and ``good"; each fuzzy event, denoted as a $2\times 2$
matrix $\widetilde{\sigma}=[a_{ij}]_{2\times 2}$, means the
possibility for patient's condition to transfer from one crisp
state to another crisp state when a certain drug treatment is
adopted. Suppose that the patient's initial condition is
$\widetilde{q}_{0}=[0.9, \hskip 1mm 0]$. The drug events
$\widetilde{a},\widetilde{b}, \widetilde{c}$, namely,
theophylline, Erythromycin Ethylsuccinate and dopamine,
respectively, may be evaluated according to doctors' experience as
follows:
\[
\widetilde{a}=\left[
\begin{array}{cc}
0.9 &\ 0.4\\
0&\ 0.4
\end{array}
\right], \widetilde{b}=\left[
\begin{array}{cc}
0.4 &\ 0.9\\
0&\ 0.4
\end{array}
\right],\widetilde{c}=\left[
\begin{array}{cc}
0.4&\ 0 \\
0.4&\ 0.9
\end{array}
\right].
\]

We specify a fuzzy set of control specifications $\widetilde{K}$
that are desired for the doctors. For the sake of simplicity, it
is assumed that $\widetilde{K}$ is ${\cal
L}_{\widetilde{G},m}$-closed. As usual, let $pr(\widetilde{K})$ be
generated by a fuzzy automaton
$\widetilde{H}=(\widetilde{Q}_{2},\widetilde{E},
\widetilde{\delta},\widetilde{p}_{0})$, where
$\widetilde{p}_{0}$=$[0.9,0]$,
$\widetilde{E}=\{\widetilde{a},\widetilde{b},\widetilde{c}\}$,
with $\widetilde{a},\widetilde{b}$ being the same as those in
$\widetilde{G}$, except that $\widetilde{c}$ is changed as
follows:
\[
\widetilde{c}=\left[
\begin{array}{cc}
0.2&\ 0 \\
0.2&\ 0.9
\end{array}
\right].
\]

For these drug events, some effects such as headache disappears
are clearly observed, but some effects may be observed only by
means of medical instruments; also, some effects such as
alleviation of pain can be controlled, but some potential side
effects may be uncontrolled. Therefore, each drug event may be
observed or controlled with some membership degrees. Suppose that
$\widetilde{\Sigma}_{uc}$ and $\widetilde{\Sigma}_{o}$ are defined
as follows:
$$\widetilde{\Sigma}_{uc}(\widetilde{a})=\widetilde{\Sigma}_{uc}(\widetilde{b})=0.1,\hskip
2mm \widetilde{\Sigma}_{uc}(\widetilde{c})= 0.2;$$$$
\widetilde{\Sigma}_{o}(\widetilde{a})=0.4,\hskip 2mm
\widetilde{\Sigma}_{o}(\widetilde{b})=0.6,\hskip 2mm
\widetilde{\Sigma}_{o}(\widetilde{c})=0.$$

In supervisory control of FDESs, the purpose of nonblocking fuzzy
supervisors is to disable the fuzzy events with respective degrees
such that the generated and marked behaviors of the supervised
system satisfy some prespecified specifications, and the
controlled system does not produce deadlocks. Therefore, for this
example, the problem is whether there exists such a nonblocking
fuzzy supervisor $\widetilde{S}_{P}:
P(\widetilde{E}^{*})\rightarrow {\cal F}(\widetilde{E})$. In the
following, we will answer the problem by proving $\widetilde{K}$
to be fuzzy controllable and fuzzy observable by means of
computing approach presented in Section IV-A.

For $\widetilde{G}$ and $\widetilde{H}$, we search for all
possible fuzzy state pairs $(\widetilde{q}_{i},\widetilde{p}_{i})$
reachable from $(\widetilde{q}_{0},\widetilde{p}_{0})$ by the
finite computing tree shown in Fig.4, which is followed by the
other three subtrees visualized by Figs. 5, 6, 7, respectively.

\vskip 5mm \setlength{\unitlength}{0.05in}

\begin{picture}(40,29)

\put(26,26){\makebox(10,5)[c]{{\tiny
$(\widetilde{q}_0,\widetilde{p}_0)=([0.9,0],$\hskip 2mm
$[0.9,0]$)}}}

\put(31,26){\line(0,-1){2}} \put(11,24){\line(1,0){40}}
\put(11,24){\vector(0,-1){5}}\put(31,24){\vector(0,-1){5}}\put(51,24){\vector(0,-1){5}}
\put(6,14){\makebox(10,5)[c]{{\tiny ($ [0.9,0.4],$\hskip 2mm
   $[0.9,0.4]$)}
}}\put(26,14){\makebox(10,5)[c]{{\tiny ($ [0.4,0.9],$\hskip 2mm
   $[0.4,0.9]$)}}}
\put(46,14){\makebox(10,5)[c]{{\tiny ($ [0.4,0],$\hskip 2mm
   $[0.2,0]$)}
}}

\put(11,19){\makebox(5,5)[c]{{\tiny
$\widetilde{a}$}}}\put(26,19){\makebox(5,5)[c]{{\tiny
$\widetilde{b}$}}} \put(51,19){\makebox(5,5)[c]{{\tiny
$\widetilde{c}$}}}

\put(11,14){\vector(0,-1){4}}

\put(31,14){\vector(0,-1){4}}

\put(51,14){\vector(0,-1){4}}

\put(8,5){\makebox(5,5)[c]{{\tiny Subtree $T_{1}
$}}}\put(29,5){\makebox(5,5)[c]{{\tiny  Subtree $ T_{2}$}}}
\put(49,5){\makebox(5,5)[c]{{\tiny Subtree $T_{3} $}}}

\put(2,0){\makebox(5,5)[l]{{\footnotesize Fig. 4. Computing tree
of all state pairs reachable from
$(\widetilde{q}_0,\widetilde{p}_0)$.} }}
\end{picture}

\setlength{\unitlength}{0.05in}
\begin{picture}(40,35)
\put(25,27){\makebox(10,5)[c]{{\tiny $([0.9,0.4],$\hskip 2mm
$[0.9,0.4]$)}}}

\put(30,28){\line(0,-1){3}} \put(10,25){\line(1,0){40}}
\put(10,25){\vector(0,-1){5}}\put(30,25){\vector(0,-1){5}}\put(50,25){\vector(0,-1){5}}
\put(5,15){\makebox(10,5)[c]{{\tiny (\underline{$
[0.9,0.4]$,\hskip 2mm
   $[0.9,0.4]$})}
}}\put(25,15){\makebox(10,5)[c]{{\tiny ($ [0.4,0.9],$\hskip 2mm
   $[0.4,0.9]$)}}}
\put(47,15){\makebox(10,5)[c]{{\tiny ($ [0.4,0.4],$\hskip 2mm
   $[0.2,0.4]$)}
}}

\put(10,20){\makebox(5,5)[c]{{\tiny
$\widetilde{a}$}}}\put(25,20){\makebox(5,5)[c]{{\tiny
$\widetilde{b}$}}} \put(50,20){\makebox(5,5)[c]{{\tiny
$\widetilde{c}$}}}

\put(30,15){\vector(0,-1){5}}

\put(50,15){\vector(0,-1){5}}

\put(27,5){\makebox(5,5)[c]{{\tiny  Subtree $ T_{2}$}}}
\put(47,5){\makebox(10,5)[c]{{\tiny (\underline{$
[0.4,0.4],$\hskip 2mm
   $[0.2,0.4]$})}
}} \put(53,10){\makebox(5,5)[c]{{\tiny $\widetilde{a}$  or
$\widetilde{b}$ or $\widetilde{c}$}}}

\put(20,0){\makebox(5,5)[l]{{\footnotesize Fig. 5. Subtree
$T_{1}$.}}}

\end{picture}

\setlength{\unitlength}{0.05in}
\begin{picture}(60,55)

\put(25,48){\makebox(10,5)[c]{{\tiny ($[0.4,0.9],$\hskip 2mm
   $[0.4,0.9])$}}}

\put(30,49){\line(0,-1){2}} \put(25,47){\line(1,0){10}}
\put(25,47){\vector(0,-1){4}}\put(35,47){\vector(0,-1){4}}

\put(13,38){\makebox(10,5)[c]{{\tiny ($ [0.4,0.4],$\hskip 2mm
   $[0.4,0.4])$}}}
\put(36,38){\makebox(10,5)[c]{{\tiny ($ [0.4,0.9],$\hskip 2mm
   $[0.2,0.9])$}}}
\put(25,38){\vector(0,-1){14}}\put(8,35){\line(1,0){17}}
\put(8,35){\vector(0,-1){5}}
\put(35,38){\vector(0,-1){14}}\put(35,35){\line(1,0){17}}
\put(52,35){\vector(0,-1){5}} \put(5,25){\makebox(10,5)[c]{{\tiny
(\underline{$ [0.4,0.4],$\hskip 2mm
   $[0.4,0.4])$}}}}
\put(14,19){\makebox(10,5)[c]{{\tiny ($ [0.4,0.4],$\hskip 2mm
   $[0.2,0.4])$}}}
\put(37,19){\makebox(10,5)[c]{{\tiny ($ [0.4,0.4],$\hskip 2mm
   $[0.2,0.4])$}}}\put(48,25){\makebox(10,5)[c]{{\tiny
(\underline{$ [0.4,0.9],$\hskip 2mm
   $[0.2,0.9])$}}}}

\put(25,19){\vector(0,-1){5}}

\put(35,19){\vector(0,-1){5}}

\put(16,9){\makebox(6,5)[c]{{\tiny (\underline{$ [0.4,0.4],$\hskip
2mm
   $[0.2,0.4])$}}}}
\put(37,9){\makebox(10,5)[c]{{\tiny \underline{($
[0.4,0.4],$\hskip 2mm
   $[0.2,0.4])$}}}}

\put(17,43){\makebox(5,5)[c]{{\tiny $\widetilde{a}$ or
$\widetilde{b}$}}}\put(25,38){\makebox(25,15)[c]{{\tiny
$\widetilde{c}$}}} \put(0,25){\makebox(25,15)[c]{{\tiny
$\widetilde{a}$ or
$\widetilde{b}$}}}\put(10,24){\makebox(25,15)[c]{{\tiny
$\widetilde{c}$}}} \put(27,24){\makebox(25,15)[c]{{\tiny
$\widetilde{a}$ or
$\widetilde{b}$}}}\put(42,25){\makebox(25,15)[c]{{\tiny
$\widetilde{c}$}}} \put(18,9){\makebox(45,15)[c]{{\tiny
$\widetilde{a}$ or $\widetilde{b}$ or $\widetilde{c}$}}}
\put(7,14){\makebox(25,5)[c]{{\tiny $\widetilde{a}$ or
$\widetilde{b}$ or $\widetilde{c}$}}}

\put(20,2){\makebox(5,5)[l]{{\footnotesize Fig. 6. Subtree
$T_{2}$.}}}

\end{picture}

\setlength{\unitlength}{0.05in}
\begin{picture}(60,33)

\put(25,28){\makebox(10,5)[c]{{\tiny ($[0.4,0],$\hskip 2mm
   $[0.2,0])$}}}

\put(30,29){\line(0,-1){3}} \put(25,26){\line(1,0){10}}
\put(25,26){\vector(0,-1){4}}\put(35,26){\vector(0,-1){4}}

\put(13,17){\makebox(10,5)[c]{{\tiny ($[0.4,0.4],$\hskip 2mm
   $[0.2,0.2])$}}}
\put(35,17){\makebox(10,5)[c]{{\tiny \underline{($ [0.4,0],$\hskip
2mm
   $[0.2,0])$}}}}
\put(25,17){\vector(0,-1){4}}
 \put(13,8){\makebox(10,5)[c]{{\tiny
(\underline{$ [0.4,0.4],$\hskip 2mm
   $[0.2,0.2])$}}}}

\put(17,22){\makebox(5,5)[c]{{\tiny $\widetilde{a}$ or
$\widetilde{b}$}}}\put(25,17){\makebox(25,15)[c]{{\tiny
$\widetilde{c}$}}} \put(18,8){\makebox(25,15)[c]{{\tiny
$\widetilde{a}$ or $\widetilde{b}$ or $\widetilde{c}$}}}

\put(20,2){\makebox(5,5)[l]{{\footnotesize Fig. 7. Subtree
$T_{3}$.}}}

\end{picture}

From above computing trees, it follows that there are only eight
different fuzzy state pairs and eight different fuzzy states
reachable from $(\widetilde{q}_{0},\widetilde{p}_{0})$ and
$\widetilde{p}_{0}$, respectively, which together with the
corresponding fuzzy event strings are listed in Table II.
Therefore, we should necessarily check the fuzzy observability
condition only when $\widetilde{s}=\epsilon$, or $\widetilde{a}$,
or $\widetilde{b}$, or $\widetilde{c}$, or
$\widetilde{b}\widetilde{a}\widetilde{c}$, or
$\widetilde{b}\widetilde{a}$, or $\widetilde{b}\widetilde{c}$, or
$\widetilde{c}\widetilde{a}$.

\begin{table}
\caption{Eight different state pairs reachable from
$(\widetilde{q}_{0},\widetilde{p}_{0})$}
\begin{center}
\begin{tabular}{|c|c|c|c|}
\hline $\widetilde{s}$  & $(\widetilde{q}_{0}\odot \widetilde{s}$,
 $\widetilde{p}_{0}\odot \widetilde{s}$)  &  $\widetilde{s}$  &
$(\widetilde{q}_{0}\odot \widetilde{s}$, $\widetilde{p}_{0}\odot \widetilde{s}$)\\
\hline $\epsilon$   & ($[0.9, 0], [0.9, 0]$)  &
$\widetilde{b}\widetilde{a}\widetilde{c}$   & ($[0.4, 0.4], [0.2, 0.4]$)\\
\hline $\widetilde{a}$   & ($[0.9, 0.4], [0.9, 0.4]$)  &
$\widetilde{b}\widetilde{a}$   & ($[0.4, 0.4], [0.4, 0.4]$)\\
\hline $\widetilde{b}$   & ($[0.4, 0.9], [0.4, 0.9]$)  &
$\widetilde{b}\widetilde{c}$   & ($[0.4, 0.9],
[0.2, 0.9]$)\\
\hline $\widetilde{c}$   & ($[0.4, 0], [0.2, 0]$)  &
$\widetilde{c}\widetilde{a}$   & ($[0.4, 0.4], [0.2, 0.2]$)\\
\hline
\end{tabular}
\end{center}
\end{table}

(1) If $\widetilde{s}=\epsilon$, from Fig. 4 we know that
$[\widetilde{p}_{0}\odot\widetilde{\sigma}]
=[\widetilde{q}_{0}\odot\widetilde{\sigma}]$ for
$\widetilde{\sigma}=\widetilde{a}$ and
$\widetilde{\sigma}=\widetilde{b}$, so the fuzzy observability
condition holds for $\widetilde{\sigma}=\widetilde{a}$ and
$\widetilde{\sigma}=\widetilde{b}$. For
$\widetilde{\sigma}=\widetilde{c}$, it is clear to know that the
fuzzy observability condition holds since
$\widetilde{D}(P(\widetilde{s}\widetilde{\sigma}))=0$.

(2) If $\widetilde{s}=\widetilde{a}$, or $\widetilde{b}$, or
$\widetilde{b}\widetilde{a}\widetilde{c}$, or
$\widetilde{b}\widetilde{a}$, or $\widetilde{b}\widetilde{c}$, we
check the fuzzy observability condition via Figs. 4, 5, 6. The
fuzzy observability condition holds obviously since in subtrees
$T_{1}$ and $T_{2}$, for any $\widetilde{s}$ and any
$\widetilde{\sigma}$, $[\widetilde{p}_{0}\odot
\widetilde{s}\odot\widetilde{\sigma}] =[\widetilde{q}_{0}\odot
\widetilde{s}\odot\widetilde{\sigma}]$.

(3) We consider the last cases of $\widetilde{s}=\widetilde{c}$,
or $\widetilde{s}=\widetilde{c}\widetilde{a}$. If
$\widetilde{s}=\widetilde{c}$,
 then $\widetilde{t}=\epsilon$, or
 $\widetilde{t}=\widetilde{c}$ such that $P(\widetilde{s})=P(\widetilde{t})$.
If $\widetilde{s}=\widetilde{c}\widetilde{a}$,
 then $\widetilde{t}=\widetilde{a}$, or
 $\widetilde{t}=\widetilde{a}\widetilde{c}$,
 or $\widetilde{t}=\widetilde{c}\widetilde{a}$
  such that $P(\widetilde{s})=P(\widetilde{t})$.
According to Fig. 7, we can test that the fuzzy observability
condition holds when
$s\in\{\widetilde{c},\widetilde{c}\widetilde{a}\}$ by means of the
following Table III.

In light of the above computing process, we have verified that
$\widetilde{K}$ satisfies the fuzzy observability condition.

On the other hand, we notice that
$\widetilde{\Sigma}_{uc}(\widetilde{\sigma})\leq 0.2$ and
$pr(\widetilde{K})(\widetilde{s}\widetilde{\sigma})\geq 0.2$ for
any $\widetilde{\sigma}\in \widetilde{E}$ and any
$\widetilde{s}\in \widetilde{E}^{*}$, so $\widetilde{K}$ satisfies
the fuzzy controllability condition clearly.

Therefore, from $\widetilde{K}$ being fuzzy observable and fuzzy
controllable together with the assumption of $\widetilde{K}$ being
${\cal L}_{\widetilde{G},m}$-closed, by Theorem 1, we know that
there exists a nonblocking fuzzy supervisor $\widetilde{S}_{P}:
P(\widetilde{E}^{*})\rightarrow {\cal F}(\widetilde{E})$ that can
disable the fuzzy events with respective degrees such that \[
{\cal L}_{\widetilde{S}_{P}/\widetilde{G}}(\widetilde{s})=
pr(\widetilde{K})^{f}(\widetilde{s}) \hskip 7mm {\rm and} \hskip
7mm {\cal L}_{\widetilde{S}_{P}/\widetilde{G},m}(\widetilde{s})=
\widetilde{K}(\widetilde{s}).
\]
In fact, $\widetilde{S}_{P}$ may be constructed as the proof of
Theorem 1 in Appendix.

\begin{table}
\caption{Testing the fuzzy observability condition for
$\widetilde{c}$ and $\widetilde{c}\widetilde{a}$}
\begin{center}
\begin {tabular}{|c|c|c|c|c|c|c|c|c|c|}
\hline
$\widetilde{s}$&$\widetilde{t}$&$\widetilde{\sigma}$&$x_{1}$&$x_{2}$&
$x_{3}$& $y$& $V$&$ W $&
$V\leq W $\\

\hline
$\widetilde{c}$&$\epsilon$&$\widetilde{a}$&0.2&0.9&0.4&0.2&0&0.08& T  \\
\cline{3-10}
&&$\widetilde{b}$&0.2&0.9&0.4&0.2&0&0.12& T \\
\cline{3-10}
 &&$\widetilde{c}$&0.2&0.2&0.4&0.2&0&0& T \\
\cline{2-10}
&$\widetilde{c}$&$\widetilde{a}$&0.2&0.2&0.4&0.2&0&0.08& T  \\
\cline{3-10}
&&$\widetilde{b}$&0.2&0.2&0.4&0.2&0&0.12& T \\
\cline{3-10}
 &&$\widetilde{c}$&0.2&0.2&0.4&0.2&0&0& T \\
\hline

$\widetilde{c}\widetilde{a}$&$\widetilde{a}$&$\widetilde{a}$&0.2&0.9&0.4&0.2&0.08&0.08& T  \\
\cline{3-10}
&&$\widetilde{b}$&0.2&0.9&0.4&0.2&0.08&0.08& T \\
\cline{3-10}
 &&$\widetilde{c}$&0.2&0.4&0.4&0.2&0.08&0.08& T \\
\cline{2-10}
&$\widetilde{a}\widetilde{c}$&$\widetilde{a}$&0.2&0.4&0.4&0.2&0.08&0.08& T  \\
\cline{3-10}
&&$\widetilde{b}$&0.2&0.4&0.4&0.2&0.08&0.08& T \\
\cline{3-10}
 &&$\widetilde{c}$&0.2&0.4&0.4&0.2&0.08&0.08& T \\
\cline{2-10}
&$\widetilde{c}\widetilde{a}$&$\widetilde{a}$&0.2&0.2&0.4&0.2&0.08&0.08& T  \\
\cline{3-10}
&&$\widetilde{b}$&0.2&0.2&0.4&0.2&0.08&0.08& T \\
\cline{3-10}
 &&$\widetilde{c}$&0.2&0.2&0.4&0.2&0.08&0.08& T \\
\hline
\end{tabular}
\end{center}
\end{table}

\section{Concluding Remarks}
Since FDES was introduced by Lin and Ying [12, 13], it has been
successfully applied to biomedical control for HIV/AIDS treatment
planning [15, 16], decision making [17] and intelligent sensory
information processing for robotic control [18, 19]. In view of
the impreciseness for some events being observable and
controllable in practice, in this paper we dealt with
Controllability and Observability Theorem, in which both the
observability and the controllability of events are considered to
be fuzzy. In particular, we have presented a computing method for
deciding whether or not the fuzzy observability and
controllability conditions hold, and thus, this can further test
the existence of supervisors in Controllability and Observability
Theorem of FDESs. As some examples (Example 3) presented show,
this computing method is clearly applied to testing the existence
of supervisors in the Controllability and Observability Theorem of
classical DESs [1], and this is a different method from classical
case [1].

As pointed out in [1], in supervisory control theory there are
three fundamental theorems: Controllability Theorem, Nonblocking
Controllability Theorem, and Controllability and Observability
Theorem. This paper, together with [20-23], has primarily
established supervisory control theory of FDESs. An further issue
is regarding the diagnosis of FDESs, as the diagnoses of classical
and probabilistic DESs [32, 33]. Also, it is worth further
considering to apply the supervisory control theory of FDESs to
practical control issues, particularly in biomedical systems and
traffic control systems [34, 35]. Moreover, dealing with FDESs
modelled by fuzzy petri nets [36] is of interest, as the issue of
DESs modelled by Petri nets [37-39].

\appendices
\section{Proof of Theorem 1}

We construct a fuzzy supervisor
$\widetilde{S}_{P}:P(\widetilde{E}^{*})\rightarrow {\cal
F}(\widetilde{E})$ as follows: $
\widetilde{S}_{P}(\epsilon)(\widetilde{\sigma})
=pr(\widetilde{K})^{f}(\widetilde{\sigma})$, and for
$\widetilde{s}\in\widetilde{E}^{*}$, \hskip 2mm
$\widetilde{S}_{P}(P(\widetilde{s}))(\widetilde{\sigma})$ is
defined by the following two cases:

{\it Case 1:} If there exists another string
$\widetilde{s^{'}}\in\widetilde{E}^{*}$ such that
$P(\widetilde{s})=P(\widetilde{s^{'}})$, then
\begin{eqnarray}
\begin{array}{ll}
\widetilde{S}_{P}(P(\widetilde{s}))(\widetilde{\sigma})=\\
\left\{
\begin{array}{ll}
\max \{\widetilde{\Sigma}_{uc}^{f}(\widetilde{\sigma}),\min\{
pr(\widetilde{K})^{f}(\widetilde{s^{'}}\widetilde{\sigma}),{\cal
L}_{\widetilde{G}}^{f}(\widetilde{s}\widetilde{\sigma})\}\},&\\
 \hskip
15mm{\rm if}\hskip 4mm
pr(\widetilde{K})(\widetilde{s}\widetilde{\sigma})\leq
pr(\widetilde{K})(\widetilde{s^{'}}\widetilde{\sigma});&\\
\max \{\widetilde{\Sigma}_{uc}^{f}(\widetilde{\sigma}),
pr(\widetilde{K})^{f}(\widetilde{s}\widetilde{\sigma})\}, & \\
\hskip 15mm{\rm if}\hskip 4mm
 pr(\widetilde{K})(\widetilde{s}\widetilde{\sigma})>
pr(\widetilde{K})(\widetilde{s^{'}}\widetilde{\sigma}).
\end{array}
\right.
\end{array}
\end{eqnarray}

{\it Case 2:} If there does not exist another string
$\widetilde{s^{'}}\in\widetilde{E}^{*}$ such that
$P(\widetilde{s})=P(\widetilde{s^{'}})$, then
\begin{equation}
\begin{array}{ll}
\widetilde{S}_{P}(P(\widetilde{s}))(\widetilde{\sigma})=\\ \left\{
\begin{array}{ll}
\min \{\widetilde{\Sigma}_{uc}^{f}(\widetilde{\sigma}), {\cal
L}_{\widetilde{G}}^{f}(\widetilde{s}\widetilde{\sigma}) \}, &{\rm
if}\ pr(\widetilde{K})(\widetilde{s}\widetilde{\sigma})\leq
\widetilde{\Sigma}_{uc}(\widetilde{\sigma}),\\
pr(\widetilde{K})^{f}(\widetilde{s}\widetilde{\sigma}), &{\rm if}\
pr(\widetilde{K})(\widetilde{s}\widetilde{\sigma})>\widetilde{\Sigma}_{uc}(\widetilde{\sigma}).
\end{array}
\right.
\end{array}
\end{equation}

Firstly we prove the sufficiency.

 1. We check the fuzzy admissibility condition. Let $\widetilde{s}\in\widetilde{E}^{*}$ and
$\widetilde{\sigma}\in \widetilde{E}$. If
$\widetilde{s}=\epsilon$, then by the fuzzy controllability
condition, we have
$$\min\{\widetilde{\Sigma}_{uc}^{f}(\widetilde{\sigma}),\hskip
1mm{\cal L}_{\widetilde{G}}(\widetilde{\sigma})\} \leq
pr(\widetilde{K})^{f}(\widetilde{\sigma})=
\widetilde{S}_{P}(\epsilon)(\widetilde{\sigma}).$$ Therefore, the
fuzzy admissibility condition holds when $\widetilde{s}=\epsilon$.
For $\widetilde{s}\neq\epsilon$, we check the fuzzy admissibility
condition by the following two cases. (i) If there exists
$\widetilde{s^{'}}\in\widetilde{E}^{*}$ such that
$P(\widetilde{s})=P(\widetilde{s^{'}})$, then from (23), we have
$$\min\{\widetilde{\Sigma}_{uc}^{f}(\widetilde{\sigma}),\hskip
1mm{\cal L}_{\widetilde{G}}(\widetilde{s}\widetilde{\sigma})\}
\leq \widetilde{\Sigma}_{uc}^{f}(\widetilde{\sigma})\leq
\widetilde{S}_{P}(P(\widetilde{s}))(\widetilde{\sigma}).$$ (ii) If
there does not exist $\widetilde{s^{'}}\in\widetilde{E}^{*}$ such
that $P(\widetilde{s})=P(\widetilde{s^{'}})$, then from (24), we
have
$$\min\{\widetilde{\Sigma}_{uc}^{f}(\widetilde{\sigma}),\hskip
1mm{\cal L}_{\widetilde{G}}(\widetilde{s}\widetilde{\sigma})\}=
\widetilde{S}_{P}(P(\widetilde{s}))(\widetilde{\sigma})$$ when
$pr(\widetilde{K})(\widetilde{s}\widetilde{\sigma})\leq
\widetilde{\Sigma}_{uc}(\widetilde{\sigma})$, and
$$\min\{\widetilde{\Sigma}_{uc}^{f}(\widetilde{\sigma}),\hskip
1mm{\cal L}_{\widetilde{G}}(\widetilde{s}\widetilde{\sigma})\}
<pr(\widetilde{K})^{f}(\widetilde{s}\widetilde{\sigma})=
\widetilde{S}_{P}(P(\widetilde{s}))(\widetilde{\sigma})$$ when
$pr(\widetilde{K})(\widetilde{s}\widetilde{\sigma})>
\widetilde{\Sigma}_{uc}(\widetilde{\sigma})$.

2. We check ${\cal
L}_{\widetilde{S}_{P}/\widetilde{G}}(\widetilde{s})=
pr(\widetilde{K})^{f}(\widetilde{s})$ for any $\widetilde{s}\in
\widetilde{E}^{*}$, where
$\widetilde{\Sigma}_{o}(\widetilde{s})>0$.  We proceed by
induction on the length of $\widetilde{s}$. If $\mid
\widetilde{s}\mid=1$, by Definition 6, $$ {\cal
L}_{\widetilde{S}_{P}/\widetilde{G}}(\widetilde{\sigma})=
\min\{{\cal L}_{\widetilde{S}_{P}/\widetilde{G}}(\epsilon), \hskip
1mm{\cal L}_{\widetilde{G}}^{f}(\widetilde{\sigma}),\hskip
1mm\widetilde{S}_{P}(\epsilon)(\widetilde{\sigma})\}.$$ Notice
that $\widetilde{S}_{P}(\epsilon)(\widetilde{\sigma})
=pr(\widetilde{K})^{f}(\widetilde{\sigma})$ and
$\widetilde{K}\subseteq {\cal L}_{\widetilde{G},m}$, we have that
${\cal L}_{\widetilde{S}_{P}/\widetilde{G}}(\widetilde{\sigma})=
pr(\widetilde{K})^{f}(\widetilde{\sigma})$. Suppose ${\cal
L}_{\widetilde{S}_{P}/\widetilde{G}}(\widetilde{s})=
pr(\widetilde{K})^{f}(\widetilde{s})$ holds for $\mid
\widetilde{s}\mid\leq k-1$ where
$\widetilde{\Sigma}_{o}(\widetilde{s})>0$. The following is to
verify the equality for any $\widetilde{s}\widetilde{\sigma}$
where $\mid \widetilde{s}\mid= k-1$. By Definition 6, and the
assumption of induction, we have
$$
{\cal
L}_{\widetilde{S}_{P}/\widetilde{G}}(\widetilde{s}\widetilde{\sigma})=
\min\{pr(K)^{f}(\widetilde{s}),\hskip
1mm{\cal L}_{G}^{f}(\widetilde{s}\widetilde{\sigma}),\hskip 1mm
S_{P}(P(\widetilde{s}))(\widetilde{\sigma})\}.
$$

Next we divide it into three cases.

(1) If there exists another string
$\widetilde{s^{'}}\in\widetilde{E}^{*}$ such that
$P(\widetilde{s})=P(\widetilde{s^{'}})$, and $
pr(\widetilde{K})(\widetilde{s}\widetilde{\sigma})\leq
pr(\widetilde{K})(\widetilde{s^{'}}\widetilde{\sigma})$, then with
the definition of
$\widetilde{S}_{P}(P(\widetilde{s}))(\widetilde{\sigma})$, we have
$${\cal
L}_{\widetilde{S}_{P}/\widetilde{G}}(\widetilde{s}\widetilde{\sigma})
=\min\{pr(K)^{f}(\widetilde{s}),\hskip 1mm {\cal
L}_{\widetilde{G}}^{f}(\widetilde{s}\widetilde{\sigma}),\hskip 1mm
\widetilde{\Sigma}_{uc}^{f}(\widetilde{\sigma})\}$$ when ${\cal
L}_{\widetilde{G}}(\widetilde{s}\widetilde{\sigma})>
pr(\widetilde{K})(\widetilde{s^{'}}\widetilde{\sigma})$ and
$\widetilde{\Sigma}_{uc}(\widetilde{\sigma})>
pr(\widetilde{K})(\widetilde{s^{'}}\widetilde{\sigma});$ and
$${\cal
L}_{\widetilde{S}_{P}/\widetilde{G}}(\widetilde{s}\widetilde{\sigma})
=\min\{pr(K)^{f}(\widetilde{s}),\hskip 1mm {\cal
L}_{\widetilde{G}}^{f}(\widetilde{s}\widetilde{\sigma}),\hskip 1mm
pr(K)^{f}(\widetilde{s^{'}}\widetilde{\sigma})\}$$ when ${\cal
L}_{\widetilde{G}}(\widetilde{s}\widetilde{\sigma})\leq
pr(\widetilde{K})(\widetilde{s^{'}}\widetilde{\sigma})$ or
$\widetilde{\Sigma}_{uc}(\widetilde{\sigma})\leq
pr(\widetilde{K})(\widetilde{s^{'}}\widetilde{\sigma})$.

By the fuzzy controllability condition and fuzzy observability
condition, we obtain that ${\cal
L}_{\widetilde{S}_{P}/\widetilde{G}}(\widetilde{s}\widetilde{\sigma})\leq
pr(\widetilde{K})^{f}(\widetilde{s}\widetilde{\sigma})$. On the
other hand, it is clear that
$pr(\widetilde{K})^{f}(\widetilde{s}\widetilde{\sigma})\leq {\cal
L}_{\widetilde{S}_{P}/\widetilde{G}}(\widetilde{s}\widetilde{\sigma})$.

(2)  If there exists another string
$\widetilde{s^{'}}\in\widetilde{E}^{*}$ such that
$P(\widetilde{s})=p(\widetilde{s^{'}})$, but $
pr(\widetilde{K})(\widetilde{s}\widetilde{\sigma})>
pr(\widetilde{K})(\widetilde{s^{'}}\widetilde{\sigma})$, then from
(23), we have
$${\cal
L}_{\widetilde{S}_{P}/\widetilde{G}}(\widetilde{s}\widetilde{\sigma})=\min\{{\cal
L}_{\widetilde{G}}^{f}(\widetilde{s}\widetilde{\sigma}),\hskip 1mm
pr(\widetilde{K})^{f}(\widetilde{s}\widetilde{\sigma})\}$$ when
$\widetilde{\Sigma}_{uc}(\widetilde{\sigma})\leq
pr(\widetilde{K})(\widetilde{s}\widetilde{\sigma})$; and
$${\cal
L}_{\widetilde{S}_{P}/\widetilde{G}}(\widetilde{s}\widetilde{\sigma})
=\min\{pr(K)^{f}(\widetilde{s}),\hskip 1mm {\cal
L}_{\widetilde{G}}^{f}(\widetilde{s}\widetilde{\sigma}),\hskip 1mm
\widetilde{\Sigma}_{uc}^{f}(\widetilde{\sigma})\}$$ when
$\widetilde{\Sigma}_{uc}(\widetilde{\sigma})>
pr(\widetilde{K})(\widetilde{s}\widetilde{\sigma})$.

Due to the fuzzy controllability condition and the assumption $
pr(\widetilde{K})(\widetilde{s}\widetilde{\sigma})>
pr(\widetilde{K})(\widetilde{s^{'}}\widetilde{\sigma})$, we have
${\cal
L}_{\widetilde{S}_{P}/\widetilde{G}}(\widetilde{s}\widetilde{\sigma})\leq
pr(\widetilde{K})^{f}(\widetilde{s}\widetilde{\sigma})$. And the
inverse ${\cal
L}_{\widetilde{S}_{P}/\widetilde{G}}(\widetilde{s}\widetilde{\sigma})\geq
pr(\widetilde{K})^{f}(\widetilde{s}\widetilde{\sigma})$ holds
clearly.

(3) If there does not exist another string $
\widetilde{s^{'}}\in\widetilde{E}^{*}$ such that
$P(\widetilde{s})=P(\widetilde{s^{'}})$, then with the definition
of $\widetilde{S}_{P}(P(\widetilde{s}))(\widetilde{\sigma})$
(i.e., Eq. (24)),  we obtain that
$${\cal
L}_{\widetilde{S}_{P}/\widetilde{G}}(\widetilde{s}\widetilde{\sigma})=
\min\{pr(K)^{f}(\widetilde{s}),\hskip 1mm {\cal
L}_{\widetilde{G}}^{f}(\widetilde{s}\widetilde{\sigma}),\hskip 1mm
\widetilde{\Sigma}_{uc}^{f}(\widetilde{\sigma})\}$$ when
$\widetilde{\Sigma}_{uc}(\widetilde{\sigma})\leq
pr(\widetilde{K})(\widetilde{s}\widetilde{\sigma})$; and
$${\cal
L}_{\widetilde{S}_{P}/\widetilde{G}}(\widetilde{s}\widetilde{\sigma})=
\min\{{\cal
L}_{\widetilde{G}}^{f}(\widetilde{s}\widetilde{\sigma}),\hskip 1mm
pr(\widetilde{K})^{f}(\widetilde{s}\widetilde{\sigma})\}$$ when
$\widetilde{\Sigma}_{uc}(\widetilde{\sigma})>
pr(\widetilde{K})(\widetilde{s}\widetilde{\sigma})$.

We can analogously verify ${\cal
L}_{\widetilde{S}_{P}/\widetilde{G}}(\widetilde{s}\widetilde{\sigma})=
pr(\widetilde{K})^{f}(\widetilde{s}\widetilde{\sigma})$ from the
fuzzy controllability condition.

3. We show that ${\cal L}_{\widetilde{S}_{P}/\widetilde{G},m}=
\widetilde{K}$ and $\widetilde{S}_{P}$ is nonblocking as follows.
Since $\widetilde{K}$ is ${\cal L}_{\widetilde{G},m}$-closed and
${\cal L}_{\widetilde{S}_{P}/\widetilde{G}}(\widetilde{s})=
pr(\widetilde{K})^{f}(\widetilde{s})$ has been proved above, by
Definition 6,
\begin{eqnarray*}
\begin{array}{lll}
&&{\cal L}_{\widetilde{S}_{P}/\widetilde{G},m}(\widetilde{s})
=\min\{{\cal L}_{\widetilde{S}_{P}/\widetilde{G}}(\widetilde{s}),
\hskip 1mm {\cal L}_{\widetilde{G},m}(\widetilde{s})\}
\\&=&\min\{pr(\widetilde{K})^{f}(\widetilde{s}), \hskip 1mm {\cal
L}_{\widetilde{G},m}(\widetilde{s})\}=\widetilde{K}(\widetilde{s}).
\end{array}
\end{eqnarray*}
Furthermore,
\begin{eqnarray*}
\begin{array}{lll}
&&\widetilde{D}(P(\widetilde{s}))\cdot pr({\cal L}_{S_{P}/
G,m})(\widetilde{s})\\&=&\widetilde{D}(P(\widetilde{s}))\cdot
pr(K)(\widetilde{s})=pr(\widetilde{K})^{f}(\widetilde{s})={\cal
L}_{S_{P}/ G}(\widetilde{s}).\end{array}
\end{eqnarray*}

We have completed the proof of {\it sufficiency}. The remainder
 is to demonstrate the {\it necessity}.

1. We prove that $\widetilde{K}$ satisfies the fuzzy
controllability condition.

Obviously, the fuzzy controllability condition holds for
$\widetilde{s}=\epsilon$. For any
$\widetilde{s}\in\widetilde{E}^{*}$, by the fuzzy admissibility
condition, we have
\begin{eqnarray*} && \min\{pr(\widetilde{K})^{f}(\widetilde{s}),\hskip 1mm
\widetilde{\Sigma}_{uc}^{f}(\widetilde{\sigma}),\hskip 1mm {\cal
L}_{G}^{f}(\widetilde{s}\widetilde{\sigma})\}\\
&\leq& \min\{{\cal
L}_{\widetilde{S}_{P}/\widetilde{G}}(\widetilde{s}),\hskip 1mm
\widetilde{S}_{P}(P(\widetilde{s}))(\widetilde{\sigma}),\hskip 1mm
{\cal L}_{\widetilde{G}}^{f}(\widetilde{s}\widetilde{\sigma} )\}\\
 &=&{\cal
L}_{\widetilde{S}_{P}/\widetilde{G}}(
\widetilde{s}\widetilde{\sigma})=pr(\widetilde{K})^{f}(\widetilde{s}\widetilde{\sigma}).
\end{eqnarray*}

 2. $\widetilde{K}$ is ${\cal L}_{\widetilde{G},m}$-closed obviously. In fact, from ${\cal
L}_{\widetilde{S}_{P}/\widetilde{G},m}=\widetilde{K}$, we have
\begin{eqnarray*}
\begin{array}{lll}&&\widetilde{K}(\widetilde{s})={\cal
L}_{\widetilde{S}_{P}/\widetilde{G},m}(\widetilde{s})\\
&=&\min\{{\cal
L}_{\widetilde{S}_{P}/\widetilde{G}}(\widetilde{s}), \hskip
1mm{\cal L}_{\widetilde{G},m}(\widetilde{s})\}\\
&=&\min\{pr(\widetilde{K})^{f}(\widetilde{s}),\hskip 1mm {\cal
L}_{\widetilde{G},m}(\widetilde{s})\}.\end{array}
\end{eqnarray*}

3. We check that $\widetilde{K}$ satisfies the fuzzy observability
condition. For any $\widetilde{s}\in\widetilde{E}^{*}$ and
$\widetilde{\sigma}\in \widetilde{E}$, if there exists another
string $\widetilde{s^{'}}\in\widetilde{E}^{*}$ such that
$P(\widetilde{s})=P(\widetilde{s^{'}})$, then the fuzzy
observability condition holds obviously if $
pr(\widetilde{K})(\widetilde{s}\widetilde{\sigma})>
pr(\widetilde{K})(\widetilde{s^{'}}\widetilde{\sigma})$. If $
pr(\widetilde{K})(\widetilde{s}\widetilde{\sigma})\leq
pr(\widetilde{K})(\widetilde{s^{'}}\widetilde{\sigma})$, we have
 \begin{eqnarray*}
&&\min\{pr(\widetilde{K})^{f}(\widetilde{s}),\hskip 2mm
pr(\widetilde{K})^{f}(\widetilde{s^{'}}\widetilde{\sigma}),\hskip
2mm {\cal
L}_{\widetilde{G}}^{f}(\widetilde{s}\widetilde{\sigma} )\}\\
&=& \min\{pr(\widetilde{K})^{f}(\widetilde{s}), \hskip 2mm{\cal
L}_{\widetilde{S}_{P}/\widetilde{G}}(\widetilde{s^{'}}\widetilde{\sigma}),\hskip
2mm
{\cal L}_{\widetilde{G}}^{f}( \widetilde{s}\widetilde{\sigma} )\}\\
&=& \min\{pr(\widetilde{K})^{f}(\widetilde{s}),\hskip 1mm{\cal
L}_{\widetilde{S}_{P}/\widetilde{G}}(\widetilde{s^{'}}),\hskip 1mm
{\cal
L}_{\widetilde{G}}^{f}(\widetilde{s^{'}}\widetilde{\sigma}),\\&&
\widetilde{S}_{P}(P(\widetilde{s^{'}})(\widetilde{\sigma}),\hskip
1mm{\cal
L}_{\widetilde{G}}^{f}( \widetilde{s}\widetilde{\sigma} )\}\\
&\leq& \min\{pr(\widetilde{K})^{f}(\widetilde{s}),\hskip 2mm {\cal
L}_{\widetilde{G}}^{f}(\widetilde{s}\widetilde{\sigma}),\hskip 2mm
\widetilde{S}_{P}(P(\widetilde{s^{'}}))(\widetilde{\sigma})\}\\
&=&{\cal
L}_{\widetilde{S}_{P}/\widetilde{G}}(\widetilde{s}\widetilde{\sigma}
)=pr(\widetilde{K})^{f}(\widetilde{s}\widetilde{\sigma}).
\end{eqnarray*}
Therefore, $\widetilde{K}$ satisfies the fuzzy observability
condition. And the proof of {\it necessity} is completed.

\section*{Acknowledgment}
The authors would like to thank the Associate Editor and the four
anonymous reviewers for their invaluable suggestions and comments
that greatly helped to improve the quality of this paper.

\ifCLASSOPTIONcaptionsoff
  \newpage
\fi



%\bibliographystyle{IEEEtran}

\end{document}